\documentclass[11pt]{article}
\usepackage[T1]{fontenc}
\usepackage[margin=1in]{geometry}
\usepackage{graphicx}
\usepackage{multirow}
\usepackage{amsmath,amssymb}
\usepackage{booktabs,tabularx}
\usepackage{subfig}
\usepackage[numbers,square,sort&compress]{natbib}
\setcitestyle{aysep={,}}
\usepackage{xurl}
\usepackage[hidelinks]{hyperref}

\newcolumntype{C}{>{\centering\arraybackslash}X}
\newcolumntype{S}{C}
\newcolumntype{L}{C}
\newcommand{\doi}[1]{\url{https://doi.org/#1}}

\title{Feature Space Novelty Detection for Open-Set Recognition of Root-Pathogen Volatiles with an Electronic Nose}
\author{Krzysztof Siwek\thanks{Corresponding author. Email: \href{mailto:krzysztof.siwek@pw.edu.pl}{krzysztof.siwek@pw.edu.pl}.},
  Tomasz Grzywacz, and Tran Hoai Linh\\
  \small Faculty of Electrical Engineering, Warsaw University of Technology\\
  \small ul. Koszykowa 75, 00-662 Warsaw, Poland}
\date{}

\begin{document}
\maketitle

\begin{abstract}
Electronic nose (e-nose) systems classify odors in agriculture, food safety, and environmental monitoring. Conventional classifiers assume a fixed set of known classes and may assign confident but incorrect labels to novel odors. We address this open-set problem with a self-training framework that combines a deep neural network, a support vector machine, and a random forest. The framework accepts a known class label only when all three classifiers agree at a conservative heuristic threshold. It routes other samples to an open-set stage that measures novelty in the learned feature space. Confirmed novel samples enter a quarantine buffer and may support the creation of a new class.
We evaluated the framework using volatiles from root pathogens cultured on germinating \textit{Quercus robur} acorns. Contiguous blocked splitting reduced leakage between adjacent autocorrelated readings, but culture-level independence could not be reconstructed. Known class accuracy and macro-averaged F1 ranged from 0.99 to 1.00. For withheld classes, the softmax confidence score approached chance, with an AUROC of 0.57. OpenMax reached 0.93, while feature space distance reached 0.97. On the Gas Sensor Array Drift Dataset, classifier fusion had a higher mean AUROC than feature space distance. The comparisons after Holm correction were not statistically significant.
We also tested the adaptive cycle. The consensus gate produced correct pseudo-labels whenever it accepted samples in the main regime. Agreement between geometric and disparity-network criteria reduced false alarms below 1\%, although sensitivity declined. Class creation was triggered in 19 of 25 simulations. Among those 19 completed updates, recognition of the new class on a segment excluded from the stream and update reached 98.2\%. The known class retention result is descriptive because its fold was reused during threshold calibration and stream simulation.
\end{abstract}


\section{Introduction}

The electronic nose (e-nose) is an artificial olfactory system that detects, identifies, and classifies volatile compounds by mimicking the mammalian sense of smell \citep{gardner1999electronic, Wilson_enos}. Over the past two decades these systems have found use in food quality monitoring, environmental control, industrial safety, and medical diagnostics \citep{Ye_review, Zhai_review}. Despite their growing importance, most deployed implementations still rely on supervised learning that assumes a fixed set of odor classes known in advance. That assumption limits scalability and robustness in the field, where previously unobserved odors can appear without warning.

Several practical settings make this limitation concrete. In agricultural disease detection, food quality control, and environmental monitoring, new odors emerge over time because of pathogen mutation, chemical contamination, or spoilage \citep{feng2022fusarium, peris2013fermentation}. In such conditions an e-nose must adapt quickly to novel classes without complete retraining or constant expert supervision, and it must keep producing reliable inferences when faced with inputs it has never seen. This motivates intelligent frameworks that learn online and reason about what they do not yet know.

Conventional machine learning models, including support vector machines (SVM) and artificial neural networks, perform well when trained on a fixed set of labeled classes \citep{loutfi2015review}. Their core weakness is the closed-world premise: every class is assumed to be available at training time, which is rarely true in practice. A softmax classifier, for example, is forced to choose among the known categories and often returns a high score even for an out-of-distribution input, so it cannot signal that a sample is unfamiliar \citep{bendale_openmax, liang_odin, hendrycks_msp}. Addressing this gap calls for a shift from static classification toward adaptive frameworks that detect and incorporate new classes as they arrive.

To make this shift, the present work draws on three complementary ideas. Open-set recognition lets a model decide whether an odor belongs to any known class rather than forcing a choice among them \citep{bendale_openmax, qu_openset}. Semi-supervised learning, and self-training in particular, expands the labeled set by assigning pseudo-labels to high-confidence unlabeled samples, which reduces the annotation burden \citep{lee_pseudolabel, chapelle2006semi}. Ensemble learning combines diverse models so that the weaknesses of any single classifier are offset by the others, improving robustness against the noise and drift typical of chemical sensors \citep{dietterich2000ensemble, rokach2010ensemble, zhou2004ensemble}.

This study integrates these ideas into a single semi-supervised adaptive framework with three central capabilities. The framework classifies odor samples into known classes using an ensemble of a deep neural network, an SVM, and a random forest; it detects novel odors by measuring the agreement and the confidence of those classifiers together with a geometric test in the learned feature space; and it updates itself by provisionally creating new classes for confirmed novel samples, so that recognition adapts continuously to evolving profiles. The contribution is therefore twofold. First, a strict three-model consensus rule is coupled to the self-training loop, so that a pseudo-label is accepted only when probabilistic, geometric, and statistical evidence point to the same class, which curtails the error propagation that ordinarily destabilizes self-training \citep{rosenberg2005learning}. Second, this consensus gate is paired with a dual-criterion novelty detector and an active quarantine-and-adapt cycle, so that the system separates ambiguous known samples from genuinely new ones before it commits to a new class. The framework is demonstrated on volatiles released by root pathogens cultured on germinating \textit{Quercus robur} acorns, a task with direct relevance to plant biosecurity and one that extends the authors' earlier e-nose work on the same biological material \citep{borowik_zoledzie, Borowik_ciboria, Jeremie_enos}.

The remainder of the paper is organized as follows. The next section reviews related work on closed-set and open-set odor recognition and on adaptive learning. The following sections summarize why fungal and oomycete detection matters for agriculture and food safety, describe the measurement hardware and the sample set, and present the classifiers and the three-stage adaptive method. The experimental protocol and an exploratory analysis of the acquired data are then reported, the strengths and limitations of the approach are discussed, and future directions are outlined.

\section{Related Work}

Research on machine learning for e-nose data has historically concentrated on closed-set problems, in which the training and test sets share the same label space. Comprehensive reviews show how feature extraction followed by SVM, random forest, or shallow neural networks became the standard pipeline, and how convolutional and recurrent networks later improved accuracy on raw or lightly processed sensor responses \citep{Ye_review, Zhai_review, Hong_array}. Within this closed-set setting, heterogeneous ensembles can improve accuracy and stability because their component models make different errors \citep{dietterich2000ensemble, mohareb2016ensemble}. These results are encouraging, yet they do not address the situation in which a test odor belongs to no training class.

The open-set problem has been formalized in the broader pattern-recognition literature. Bendale and Boult introduced the OpenMax layer, which replaces the closed-set softmax with an estimate of the probability that an input belongs to none of the known classes, using the distance between an activation vector and per-class mean vectors together with extreme-value statistics \citep{bendale_openmax}. Liang and co-workers proposed ODIN, which sharpens the separation between in-distribution and out-of-distribution inputs through temperature scaling and small input perturbations, and which requires almost no change to a trained network \citep{liang_odin}. These methods underpin a family of distance-based and prototype-based detectors in which a sample is compared with class centroids in a learned embedding, and they directly inspire the geometric novelty test used in the present work.

For gas sensing specifically, open-set recognition has only recently received sustained attention. Qu and colleagues studied open-set learning on a public e-nose dataset and showed that classifiers built on a robust latent space can reject unseen substances while still recognizing the known ones, which establishes the feasibility of the approach for sensor arrays \citep{qu_openset}. Parallel efforts treat the appearance of unknown gases jointly with sensor drift, since both push the test distribution away from the training distribution, and they argue that the two problems are best handled within a single open-set risk formulation. These studies confirm that softmax thresholding alone is insufficient and that an explicit rejection mechanism is needed for reliable field operation.

Reducing the dependence on labeled data is the second strand of relevant work. Self-training assigns pseudo-labels to confident unlabeled samples and folds them back into the training set \citep{lee_pseudolabel}. The method is simple and effective, but it is also fragile, because an early mistake can be reinforced over successive iterations and degrade the model \citep{rosenberg2005learning}. The standard remedies tighten the acceptance criterion or introduce diversity, for instance by requiring agreement among several models before a pseudo-label is trusted \citep{zhou2004ensemble, chapelle2006semi}. The consensus rule adopted here follows this logic and makes the agreement explicit across three classifiers with distinct inductive biases.

A third strand concerns rapid adaptation when only a few examples of a new class are available. Meta-learning trains a model to adapt quickly to unseen tasks, with Model-Agnostic Meta-Learning optimizing an initialization from which a few gradient steps suffice \citep{ICML_Finn}, and prototypical networks classifying by proximity to class prototypes in an embedding space \citep{Snell_NeurIPS}. Such few-shot ideas are attractive for olfaction, where labeled samples of a newly encountered odor are scarce, and they have been linked to continual-learning schemes that resist sensor drift over time \citep{liu_deepsniffer}. A recent survey of few-shot action recognition, although centered on video, offers a taxonomy of generative and meta-learning approaches that transfers naturally to odor learning, in particular to instance representation, prototype learning, and generalized matching \citep{Wanyan_few}. Reinforcement and active learning complete the picture, since an agent can update its model from the outcome of its own actions and can query a human only for the most informative samples, which is valuable when annotation is costly. A study that applied deep reinforcement learning to ventilation control in pig housing, with the aim of reducing ammonia and odor while saving energy, illustrates how sensor-driven control can be optimized even outside a classical e-nose, and its findings transfer to odor-control systems built around sensor arrays \citep{Zheng_reinf}.

Despite this progress, few studies combine self-training with dynamic class expansion in an e-nose, particularly under the practical constraints of field deployment. The framework proposed here is positioned in exactly that gap: it keeps the strong closed-set accuracy of an ensemble, adds an explicit open-set rejection stage, and closes the loop with an adaptation cycle that creates new classes only after careful verification.

\section{Significance of Fungal and Oomycete Detection}

The detection of fungal and oomycete pathogens in agricultural products is a critical task for food safety and economic stability. The impact of these pathogens is twofold. They reduce crop yield, lowering the quantity of the harvest, and they degrade its quality by contaminating grain with mycotoxins, the toxic secondary metabolites produced during fungal growth.

Fungi of the genus \textit{Fusarium} pose a dual threat, acting both as human pathogens that infect the eyes and fingernails and as devastating agricultural pests. The genus is ubiquitous and adapts to many ecological niches and building materials, and among its members \textit{Fusarium graminearum} and \textit{Fusarium oxysporum} are ranked internationally among the ten most significant plant pathogens \citep{dean_top10fungal}. The risk associated with \textit{Fusarium} remains high after harvest, because grain stored in silos provides conditions that favor fungal growth, rendering the grain unfit for consumption and creating a need for rapid tools that identify spoilage in collected samples.

Beyond \textit{Fusarium}, several other soil-borne and water-borne pathogens threaten global agriculture and plant health, in particular \textit{Phytophthora}, \textit{Pythium}, \textit{Rhizoctonia}, and \textit{Cylindrocarpon}. \textit{Phytophthora} and \textit{Pythium} are oomycetes, often called water molds, and are notorious for late blight, damping-off, and aggressive root rot across a wide range of crops and ornamentals. \textit{Rhizoctonia} is a true fungal genus that causes sore shin, stem cankers, and brown patch, and that survives in soil for long periods through resilient sclerotia. \textit{Cylindrocarpon} species are associated with root-rot complexes and black-foot disease, especially in perennial crops such as grapevines and fruit trees, where they cause progressive necrosis and systemic decline. As with \textit{Fusarium}, these pathogens are economically damaging, and their early detection is central to biosecurity and productivity.

Traditional diagnostic methods such as classical cultivation, immunological assays, and genetic sequencing by the polymerase chain reaction offer high sensitivity and accuracy, but they are constrained by high cost and the need for specialized laboratory conditions. A practical alternative is to detect the volatile organic compounds (VOCs) released into the surrounding air during fungal growth, which provide a chemical signature of infestation \citep{Jeremie_enos}. This signature can be read by an electronic nose, a device that uses an array of non-specific gas sensors with overlapping detection ranges \citep{Wilson_enos}. Unlike gas chromatography coupled with mass spectrometry, which identifies individual chemical components, the e-nose relies on pattern recognition over the collective response of the array \citep{Borowik_enos1}. By analyzing that response, such systems classify samples and distinguish healthy from infected grain with accuracies that earlier studies on related material reported to exceed eighty percent \citep{borowik_zoledzie, Jeremie_enos}. These high-throughput analyzers are fast, non-destructive, and cost-effective, which makes them well suited to real-time monitoring of grain safety in large storage facilities.

\begin{figure}[htbp]
    \centering
    \includegraphics[width=0.85\linewidth]{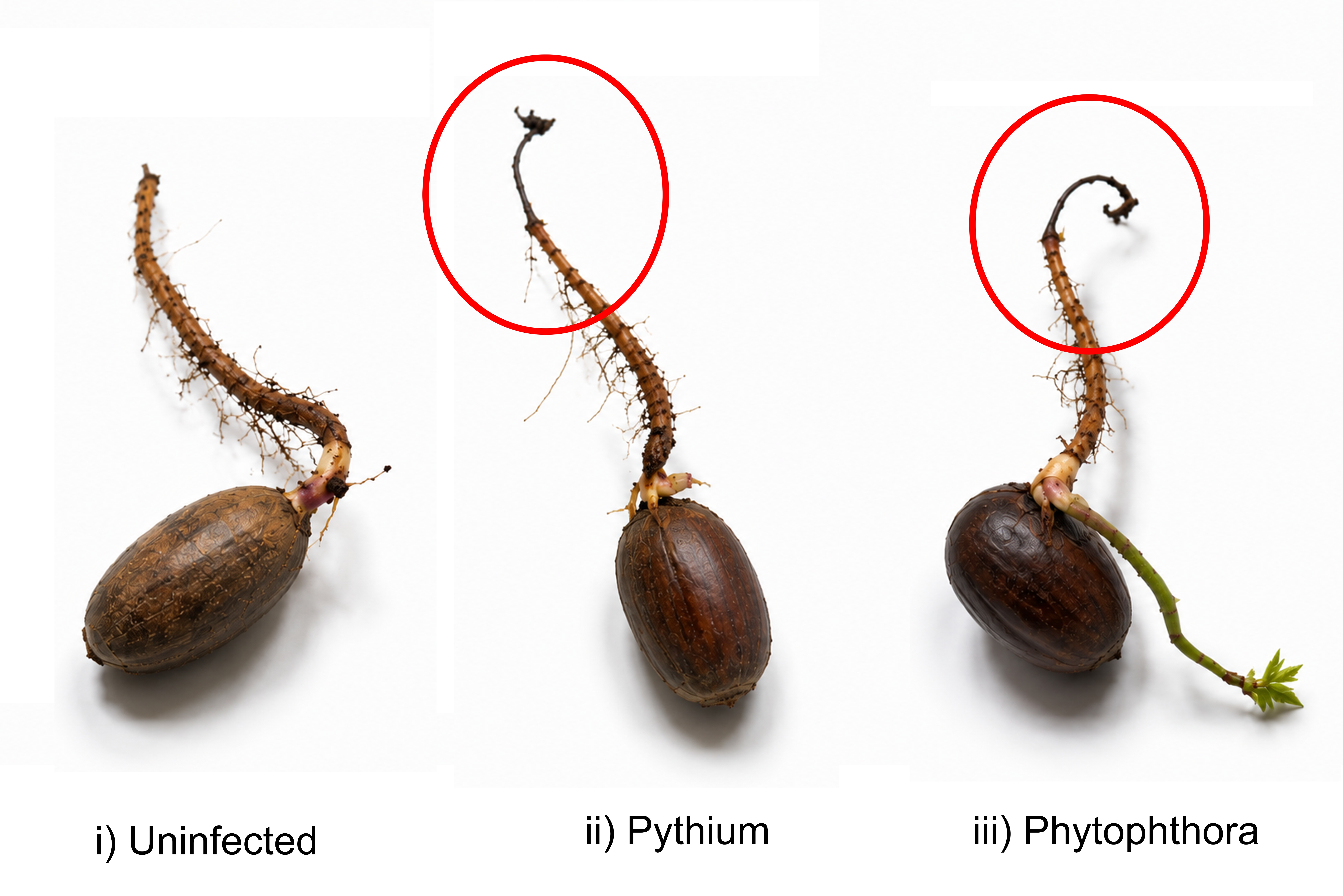}
    \caption{Germinating \textit{Quercus robur} acorn seedlings used in the study: i) healthy control seedling;  ii) seedling infected by \textit{Pythium}; iii) seedling infected by \textit{Phytophthora}. Red ellipses mark necrotic lesions on the radicle.}
    \label{fig:acorns}
\end{figure}

\section{Materials and Methods}

\subsection{Problem Formulation}
\label{sec:problem}

The task is stated formally before the system is described. Let $\mathcal{K} = \{1, \dots, n_K\}$ be the set of known odor classes seen during training, with $n_K = |\mathcal{K}|$ their number, and let $\mathcal{D} = \{(\mathbf{x}_i, y_i)\}_{i=1}^{N}$, with $\mathbf{x}_i \in \mathbb{R}^{d}$ and $y_i \in \mathcal{K}$, be the labeled training set, where $d$ is the number of sensor features and $N$ is the number of training samples. At deployment a sample may belong to a known class or to a class never seen before, so the target is an open-set classifier $g : \mathbb{R}^{d} \rightarrow \mathcal{K} \cup \{\text{unknown}\}$ rather than the closed-set map $f : \mathbb{R}^{d} \rightarrow \mathcal{K}$ that a conventional model learns. The classifier is built from a closed-set predictor that supplies class posteriors $p(k \mid \mathbf{x})$ and from a novelty score $s(\mathbf{x}) \in \mathbb{R}$ that is large when the sample is unfamiliar, and a sample is referred to the unknown route when $s(\mathbf{x})$ exceeds a threshold $\tau$, otherwise it is assigned to $\arg\max_{k} p(k \mid \mathbf{x})$. The central design question of this paper is how the novelty score $s$ should be defined, since it can be read from the output probabilities or from the geometry of an internal representation, and these choices behave very differently. Performance is therefore not a single accuracy but a pair of competing quantities, the closed-set error on known samples and the ability to separate known from unknown, and the objective is to choose the model and the score that jointly minimize the closed-set error while maximizing the threshold-free separation of knowns and unknowns, the latter summarized by the area under the receiver operating characteristic curve of $s$ and by the normalized accuracy that weights correct known classification and correct unknown rejection equally.

\subsection{System Architecture}

\begin{figure}[htbp]
    \centering
    \includegraphics[width=0.85\columnwidth]{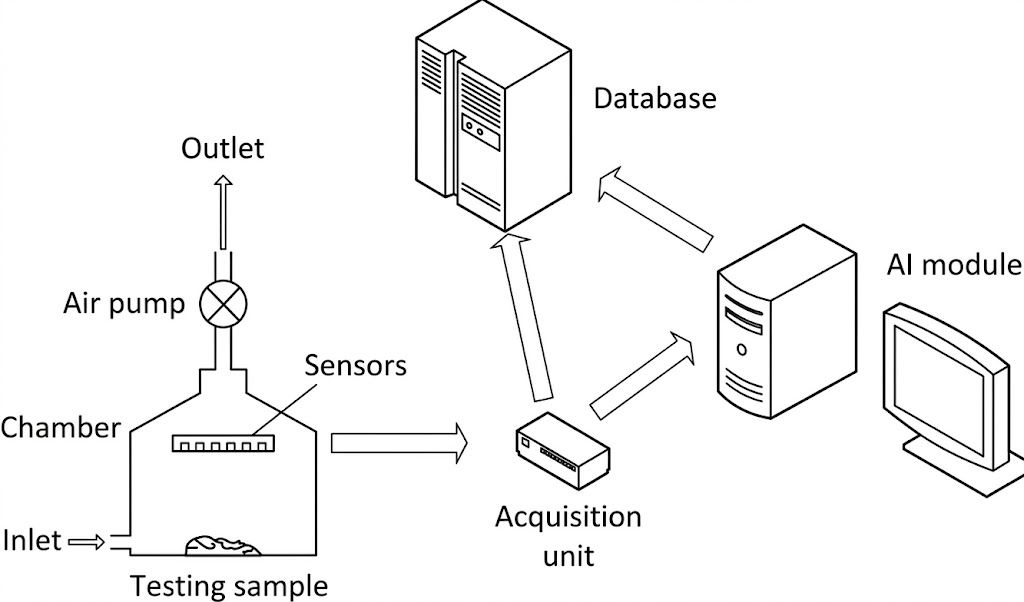}
    \caption{Schematic diagram of the electronic nose, comprising the pneumatic sample-delivery path, the sensor chamber with the array, the data-acquisition and preprocessing electronics, and the host computer that runs the classification and adaptation pipeline.}
    \label{fig:enos_schema}
\end{figure}

The effectiveness of an e-nose depends not only on its learning algorithms but also on the physical design of the hardware that senses the VOCs. The system used in this study comprises four functional parts that work together. A sensor chamber fitted with a deflector ensures consistent airflow and a uniform distribution of the odor across the array. A signal-conditioning and data-acquisition module preprocesses the sensor responses and transmits them to a computer for analysis. A classification stage hosts the three models, a deep multilayer neural network, a support vector machine, and a random forest, each producing an independent prediction. Finally, an adaptive learning mechanism updates the system when a new class is identified. Figure~\ref{fig:enos_schema} shows the overall architecture.

\subsection{E-Nose Hardware Implementation}

The sensor technologies most widely used in e-nose systems include metal-oxide semiconductors (MOS), conducting polymers, quartz crystal microbalances (QCM), surface acoustic wave (SAW) devices, and, more recently, carbon-nanotube and graphene-based materials \citep{Wilson_enos, Zhai_review}. Each class differs in response time, recovery kinetics, baseline stability, and cross-sensitivity. MOS sensors are favored for their low cost and high sensitivity to a broad range of gases, although they are prone to baseline drift and humidity interference. By contrast, QCM and SAW sensors offer high mass sensitivity and precision, but they often require complex signal processing and careful environmental control to suppress thermal noise.

Complementing these established technologies, complementary metal-oxide-semiconductor (CMOS) gas sensors have gained traction because of their potential for dense integration, miniaturisation, and low power consumption. CMOS technology allows the sensing element to be integrated monolithically with the analogue front end and the digital processing on a single chip, which suits portable and wearable applications. Such sensors typically use micro-hotplates embedded in the CMOS substrate to regulate the temperature of the gas-sensitive material, and the resulting temperature-modulation protocols can extract additional kinetic features from the analyte interaction, improving selectivity toward VOCs. Advances in CMOS-compatible nanomaterials, including metal-oxide nanowires and two-dimensional materials, have further raised sensitivity while remaining compatible with standard semiconductor fabrication \citep{Trung_CMOS}.

\begin{figure}[htbp]
    \centering
    \includegraphics[width=0.8\linewidth]{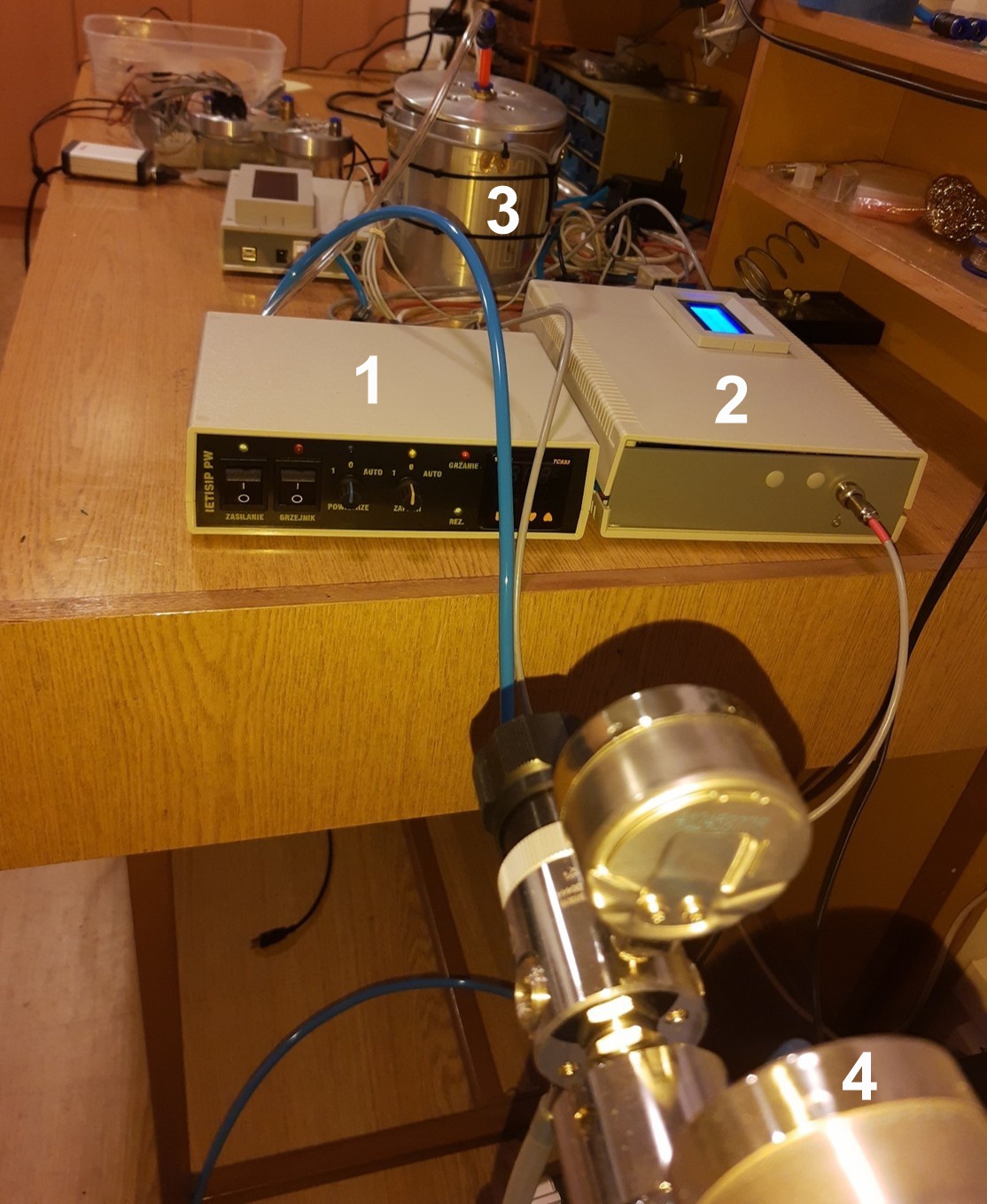}
    \caption{Laboratory realisation of the constructed electronic nose, showing the control and heater unit, the data-acquisition module, and the gold-plated sensor chamber connected through the pneumatic path.}
    \label{fig:meas_system}
\end{figure}

The architecture in Figure~\ref{fig:enos_schema} contains three subsystems. The sample delivery system draws vapor into the sensor chamber and stabilizes the flow. The detection unit contains non-specific, cross-sensitive sensors \citep{sto_enos_brudz}. In MOS devices, analytes react with chemisorbed oxygen, thereby changing the measurable resistance. The acquisition unit converts the analog responses to digital values. For the present study, processing consisted solely of the baseline normalization defined in Section~\ref{sec:preprocessing}. We applied no software filtering, smoothing, or transient feature extraction. Figure~\ref{fig:meas_system} shows the laboratory system.

State-of-the-art commercial platforms, such as the PEN3 by AIRSENSE: \citep{pen3} and the Cyranose 320 \citep{Cyranose_320}, standardized modular arrays combined with embedded pattern recognition. The current trend, however, is toward miniaturised and wearable architectures that use micro-electromechanical (MEMS) sensors and wireless communication for distributed sensing in healthcare and environmental monitoring \citep{Rock_enos, wilson2011applications}. These developments supply the high-dimensional data streams that the open-set methods proposed here require.

\subsection{Sample Preparation and Signal Acquisition}
\label{sec:preprocessing}

\begin{figure}[htbp]
    \centering
    \includegraphics[width=0.8\linewidth]{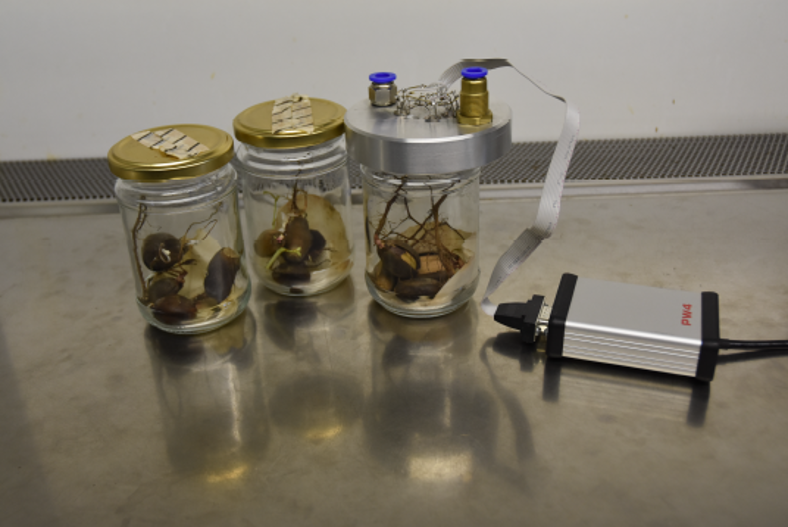}
    \caption{Acorn samples sealed in glass jars and connected to the portable acquisition unit, which transfers the accumulated headspace from each jar to the sensor chamber of the electronic nose.}
    \label{fig:meas_acorn}
\end{figure}

\begingroup
We used the constructed e-nose to analyze VOCs emitted by parasitic fungi and oomycetes associated with tree roots. Germinating pedunculate oak acorns (\textit{Quercus robur} L.) served as the biological substrate \citep{borowik_zoledzie, Borowik_enos1, Jeremie_enos, Borowik_enos2}. We studied \textit{Cylindrocarpon}, \textit{Fusarium}, \textit{Rhizoctonia}, \textit{Pythium}, and \textit{Phytophthora}. We cultivated each pathogen separately on an artificial growth medium. We incubated the acorns in sterilized moist sand under light at $25\,^{\circ}$C. After one month, we placed acorn tips with radicles of 5--7 cm on agar plugs containing mycelium. We used separate Petri dishes for each experimental variant. We moistened the filter paper in each dish with 5 mL of sterile distilled water. We then incubated the dishes under daylight at approximately $20$--$22\,^{\circ}$C. We inspected them every 8 h until seedlings collapsed and all roots showed necrosis. We transferred the acorns into measurement jars and reisolated the pathogens in accordance with Koch's postulates.

We prepared five independent cultures on five consecutive days for each pathogen and the uninfected medium control. Each culture contributed three jars, giving fifteen jars per class. The 330 readings per class therefore averaged 22 readings per jar. These readings were closely spaced observations from the same headspace exposures, not independent biological replicates. The pump delivered headspace at 100 mL/min, and the system recorded each sensor channel every 1.2 s. We allowed the sensors to recover for 5 min under ambient-air flow after each measurement. Figure~\ref{fig:meas_acorn} shows the sealed jars connected to the sampling unit. The sensing chamber contained fourteen Figaro channels, two each of TGS2600, TGS2602, TGS2603, TGS2610, TGS2611, TGS2616, and TGS2620, operated at two sensitivity points.

We applied baseline division but no software filtering, smoothing, or noise-reduction procedure. Let $v_j(t)$ denote the recorded output of sensor channel $j$ at reading $t$. We defined its clean air baseline as

\begin{equation}
 b_j=\frac{1}{100}\sum_{t=1}^{100}v_j(t),
 \label{eq:baseline}
\end{equation}

and the normalized response as

\begin{equation}
 x_j(t)=\frac{v_j(t)}{b_j}.
 \label{eq:normalised_response}
\end{equation}

We treated readings 300--400, after the response had stabilized, as steady-state observations. Each model input was the 14 dimensional vector $\mathbf{x}(t)=[x_1(t),\ldots,x_{14}(t)]^{\top}$ from that interval.
\endgroup

\begin{figure}[htbp]
    \centering
    \includegraphics[width=0.85\columnwidth]{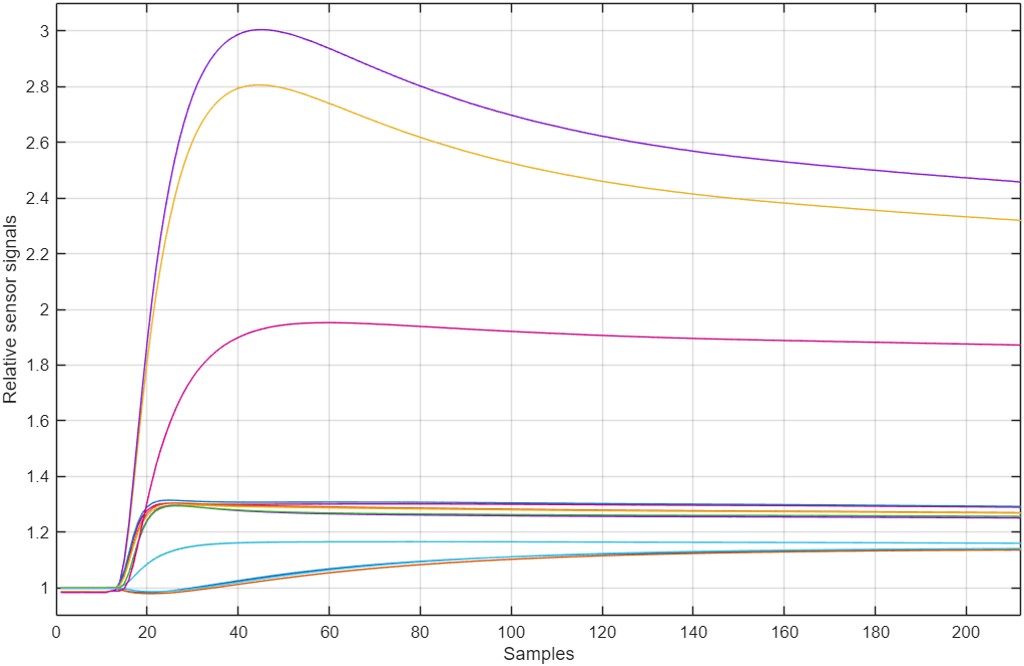}
    \caption{Measured responses of the Figaro MOS sensors of the constructed electronic nose during a single measurement cycle.}
    \label{fig:signal_sensors}
\end{figure}

The system logged all sensor outputs to a structured database. Figure~\ref{fig:signal_sensors} shows one measurement cycle. The analysis used the baseline normalized steady-state vectors defined in Equations~\eqref{eq:baseline} and~\eqref{eq:normalised_response}. It did not apply smoothing, digital noise reduction, or transient feature extraction. We used the same vectors for classification and exploratory clustering.

\subsection{Base Classifiers}

The classification ensemble combines three models with deliberately different inductive biases, so that a sample is treated as known only when probabilistic, geometric, and statistical evidence agree.

The deep neural network (DNN) is the probabilistic and non-linear pillar of the ensemble, designed to interpret the multi-dimensional patterns produced by the array during exposure. It narrows to a low-dimensional bottleneck layer placed immediately before its output, which serves both as a compact representation for classification and as the feature space in which the open-set novelty distance is later measured. Because the two datasets studied here differ in input dimension, the network takes two related forms, a convolutional model for the fourteen-channel acorn signals and a fully connected model for the higher-dimensional drift descriptors, and both are specified in full in Section~\ref{sec:nn}.

The support vector machine (SVM) is the geometric pillar of the ensemble, chosen for its strong generalization in high-dimensional feature spaces \citep{Cortes_svm}. Equation~\eqref{eq:svm_margin} states the classical two-class, hard-margin formulation, shown here for its clarity as the starting point of the method rather than as a description of what is fitted, with $w$ the normal vector of the separating hyperplane, $b$ its offset, and $(\mathbf{x}_i, y_i)$ a training pair carrying the binary label $y_i \in \{-1, +1\}$ this classical formulation assumes, distinct from the multiclass label $y_i \in \mathcal{K}$ defined in Section~\ref{sec:problem}. It identifies the hyperplane that separates the odor classes with the largest possible margin, as expressed in Equation~\eqref{eq:svm_margin}. By maximizing the margin, the SVM becomes less sensitive to noise and to minor drift in the sensor signals, which improves reliability in the field.

\begin{equation}
    \min_{w, b} \tfrac{1}{2}\,\lVert w \rVert^{2}
    \quad \text{subject to} \quad
    y_i\,(w \cdot x_i + b) \geq 1
    \label{eq:svm_margin}
\end{equation}

The classifier actually fitted, and the one every reported result uses, differs from Equation~\eqref{eq:svm_margin} in two respects. It is a soft-margin formulation, in which the regularisation constant $C$ given in Table~\ref{tab:repro} allows a controlled number of margin violations rather than forbidding them outright, and it is extended to the $n_K$ known classes of the leave-one-class-out split by scikit-learn's standard one-versus-one reduction rather than the two-class case shown above. Because the chemiresistive responses are not linearly separable, the SVM uses the radial basis function (RBF) kernel in Equation~\eqref{eq:svm_rbf}, which maps the inputs into a high-dimensional space where complex odor patterns can be separated linearly.

\begin{equation}
    \kappa(x_i, x_j) = \exp\!\left(-\gamma\,\lVert x_i - x_j \rVert^{2}\right)
    \label{eq:svm_rbf}
\end{equation}

Here $x_i$ and $x_j$ are any two feature vectors being compared, not necessarily training points, and $\gamma > 0$ is the kernel bandwidth, whose value for each dataset is given in Table~\ref{tab:repro}. The random forest (RF) is the statistical pillar, providing a robust non-parametric model that copes well with the high-dimensional and noisy data produced by MOS arrays \citep{Breiman_rf}. It consists of $M$ decision trees, each grown on a bootstrap sample of the training data, with a random subset of the sensor features considered at every split. Writing $T_m(x)$ for the class predicted by the $m$-th tree on sample $x$, the forest classifies a sample $x$ by majority vote over the trees, as in Equation~\eqref{eq:rf}. This bagging reduces the variance of the model and makes it resilient to sensor-specific noise and to outliers that could mislead a single tree.

\begin{equation}
    \hat{y} = \operatorname{mode}\{\,T_1(x), T_2(x), \dots, T_M(x)\,\}
    \label{eq:rf}
\end{equation}

Here $\hat y$ is the forest's predicted label. The three models therefore offer distinct and complementary views. The DNN contributes a probabilistic confidence through the Softmax output, the SVM contributes a boundary-based confidence that is high only when a sample lies far from the decision surface, and the RF contributes a vote-based confidence that grows with the fraction of trees that agree. In the self-training loop these views are combined so that an unlabeled sample is accepted as known only when all three reach a high-confidence agreement, which provides a triple-checked gate for the updates described next.

\subsection{Neural Network Architectures}
\label{sec:nn}

Both datasets are classified by a compact feedforward network trained from scratch on each cross-validation fold, and the two networks share a deliberate design choice that the open-set stage depends on. Each network narrows to a low-dimensional bottleneck layer placed directly before the linear output, so that every odor is compressed into a few coordinates, and it is this bottleneck representation, rather than the softmax output, in which the Mahalanobis novelty distance of the open-set stage is measured. Hidden units use the rectified linear unit, and the output layer applies the softmax function $\sigma$ of Equation~\eqref{eq:softmax} to the logit vector $\mathbf{z} = (z_1, \dots, z_{n_K})$ it produces, turning the final scores into class probabilities over the $n_K$ known classes of that fold. Training minimises the categorical cross-entropy loss with the Adam optimizer \citep{Kingma_adam}, which maintains exponentially decaying averages of past gradients and past squared gradients to give per-parameter adaptive step sizes, and which converges quickly and stays stable in the presence of the drift and noise characteristic of metal-oxide sensors.

\begin{equation}
\sigma(\mathbf{z})_i = \frac{e^{z_i}}{\sum_{j=1}^{n_K} e^{z_j}}, \qquad i = 1, \dots, n_K
\label{eq:softmax}
\end{equation}

For the acorn data the fourteen stabilised sensor responses are treated as a single channel of length fourteen and passed to a one-dimensional convolutional network. A convolution with eight filters of width three slides across neighbouring sensors, so each filter learns a local response pattern shared along the array, and after a rectified-linear activation the eight resulting feature maps, each of length twelve, are flattened into a ninety-six dimensional vector. A fully connected layer reduces this to a five-unit bottleneck, and a final linear layer maps the bottleneck to the logits of the five known classes, since one of the six classes is withheld in every run. The network is trained for four hundred epochs at a learning rate of 0.01 with a weight decay of $10^{-4}$ and He initialization, and the five-unit bottleneck is the feature space used by the distance-based novelty detector.

For the drift benchmark each measurement is already a flat descriptor vector of 128 values, eight summary features for each of the sixteen sensors, so the ordered-channel structure that the convolution exploits is absent and a fully connected network is the natural choice. A first layer maps the 128 inputs to thirty-two units with a rectified-linear activation, a second layer reduces these to an eight-unit bottleneck, and a linear output layer produces the five known-class logits. It is trained for eighty epochs at a learning rate of 0.005 with the same weight decay, and the eight-unit bottleneck plays exactly the role of the five-unit bottleneck in the convolutional model. The two architectures and their training settings are collected in Table~\ref{tab:nn}.

\begin{table}[t]
\caption{Architecture and training settings of the two neural networks. Both end in a narrow bottleneck placed before the output layer, which is the feature space used by the distance-based open-set detector. Each run has five known classes because one of the six is withheld as the unknown.}
\label{tab:nn}
\begin{tabularx}{\textwidth}{lXX}
\toprule
 & \textbf{Acorn network} & \textbf{Drift benchmark network} \\
\midrule
Input & 14 sensor channels (length-14 signal) & 128 descriptors (8 per sensor) \\
Type & one-dimensional convolutional & fully connected (MLP) \\
Feature stage & Conv1D, 8 filters of width 3, ReLU, flattened to 96 & FC $128\to32$, ReLU \\
Bottleneck & FC $96\to5$, linear & FC $32\to8$, linear \\
Output & FC $5\to5$ classes, softmax & FC $8\to5$ classes, softmax \\
Loss & categorical cross-entropy & categorical cross-entropy \\
Optimiser & Adam & Adam \\
Learning rate & 0.01 & 0.005 \\
Epochs & 400 & 80 \\
Weight decay & $10^{-4}$ & $10^{-4}$ \\
Initialisation & He & He \\
\bottomrule
\end{tabularx}
\end{table}

All models were implemented in Python with numpy and scikit-learn, and every result uses a fixed random seed for full reproducibility. The hyperparameters were set to standard values and were deliberately not tuned on the test data, so as to avoid optimistic bias. The support vector machine uses a radial basis kernel with $C = 10$ for the fourteen-channel acorn data and a linear kernel for the higher-dimensional drift data, the random forest uses three hundred trees for the acorn data and one hundred for the drift data, and the $k$-nearest-neighbor baseline uses ten distance-weighted neighbors, while the two networks use the settings listed in Table~\ref{tab:nn}. Table~\ref{tab:repro} collects the random seed, the exact software versions, and the remaining implementation details, the SVM kernel parameters and calibration method, the random forest and $k$-nearest-neighbor settings, and the configuration of OpenMax, temperature scaling, and the energy score, for every method compared in Table~\ref{tab:results}. Feature standardisation is fitted on the training fold alone and then applied to the test fold, so that no information from the test fold enters either training or threshold selection. Because the leave-one-class-out protocol trains a fresh model for every combination of withheld class and fold, and different combinations share most of their known-class training data, whether two combinations differ only in which fold is held out or in which class is withheld, these combinations are not statistically independent of one another. Wherever this paper reports a confidence interval or a significance test, the withheld class, the pathogen on the acorn data or the gas on the drift benchmark, rather than the class-by-fold combination, is treated as the independent statistical unit, and results within a withheld class are averaged before any interval or test is computed. The complete code that regenerates every table and figure is provided in the accompanying scripts.

\begin{table}[t]
\caption{Reproducibility summary. Every experiment uses the fixed seed and library versions below, and no setting listed here was tuned on test data. Network architectures and their training settings are given separately in Table~\ref{tab:nn}, and the disparity network's architecture, training, and boundary-perturbed data construction are given in Section~\ref{sec:hybrid}.}
\label{tab:repro}
\begin{tabularx}{\textwidth}{lX}
\toprule
\textbf{Setting} & \textbf{Value} \\
\midrule
Random seed & $20260614$, used for every model, every fold, and every bootstrap or permutation procedure in this paper \\
Software & Python 3.12, NumPy 2.4, SciPy 1.17, scikit-learn 1.8 \\
DNN implementation & Implemented directly in NumPy, forward pass, backpropagation, and the Adam optimizer all hand-written; no deep-learning framework is used \\
Adam optimizer, all three networks & $\beta_1 = 0.9$, $\beta_2 = 0.999$, $\epsilon = 10^{-8}$; per-network learning rate, epochs, and weight decay are given in Table~\ref{tab:nn} and Section~\ref{sec:hybrid} \\
SVM, acorn data & scikit-learn \texttt{SVC}, RBF kernel, $C = 10$, $\gamma = \texttt{'scale'}$, which resolves to $1/(n_{\text{features}} \cdot \mathrm{Var}(X)) \approx 0.071$ on the standardized fourteen-feature data; \texttt{probability=True}, which fits a Platt-scaling sigmoid using libsvm's built-in five-fold internal cross-validation \citep{lin_platt_svm} \\
SVM, drift benchmark & scikit-learn \texttt{LinearSVC}, $C = 0.5$, up to $2000$ iterations, wrapped in \texttt{CalibratedClassifierCV} with three-fold internal cross-validation for Platt-style probability estimates \citep{lin_platt_svm} \\
Random forest, acorn data & $300$ trees, \texttt{max\_features='sqrt'}, all other settings at scikit-learn defaults \\
Random forest, drift benchmark & $100$ trees, \texttt{max\_features='sqrt'}, all other settings at scikit-learn defaults \\
$k$-nearest neighbors, acorn data only & $k = 10$, distance-weighted voting \\
OpenMax, acorn data only & Weibull tail fitted on the twenty largest per-class activation distances, location fixed at zero, scale floored at $10^{-6}$, recalibration parameter $\alpha$ equal to the number of known classes \citep{bendale_openmax} \\
Temperature scaling, acorn data only & Single scalar temperature fitted by minimising negative log-likelihood on the training logits with a bounded scalar optimizer, searched over $T \in [0.05, 50]$ \citep{guo_calibration} \\
Energy score, acorn data only & $-\mathrm{logsumexp}(\text{logits})$, no free parameters \citep{liu_energy} \\
\bottomrule
\end{tabularx}
\end{table}

\subsection{Consensus-Based Pseudo-Labelling and Self-Training}
\label{sec:consensus}

The adaptive method transforms the e-nose from a static classifier into an open-set system that can improve itself, and it operates in three connected stages: consensus-based classification with self-training, hybrid novelty detection, and active adaptation.

The first stage maximises accuracy on known analytes while exploiting unlabeled data to refine the decision boundaries. The primary classification is performed by the ensemble of the DNN, the SVM, and the RF, whose diversity limits the bias of any one model and improves robustness against noise and drift. A new sample $x$ is assigned to a known class $k$ only when a strict consensus is reached, meaning that all three models predict the same class with a confidence that exceeds a threshold, as stated in Equation~\eqref{eq:consensus}.

\begin{equation}
    \operatorname{conf}(k \mid x) \geq \theta_{C}
    \quad \text{for every model in the ensemble}
    \label{eq:consensus}
\end{equation}

We use $\theta_C=0.90$ as a conservative heuristic operating point for the unanimity gate, not as a generally calibrated confidence threshold. We also test $\theta_C=0.80$ in Section~\ref{sec:adaptive_validation}. The three scores do not share a common probabilistic calibration. The DNN supplies its raw maximum softmax value. The SVM supplies a Platt-scaled probability fitted within the training fold \citep{lin_platt_svm}. The random forest supplies the uncalibrated fraction of trees voting for its predicted class. We apply the same numerical cutoff only to impose a simple, reproducible rule: all predictions must agree, and the minimum score must exceed $\theta_C$. We therefore interpret the acceptance rate and pseudo-label precision at this operating point, rather than the score itself as a calibrated probability. Section~\ref{sec:adaptive_validation} shows that acceptance depends on seed size and the chosen cutoff. When the gate accepts an unlabeled sample, we add its pseudo-label to the next training round \citep{lee_pseudolabel}. Otherwise, we route the sample to the novelty stage.

Two distinct combinations of the ensemble are used in this work and should not be confused. The consensus gate just described is a hard unanimity rule, in which a known-class label is accepted only when all classifiers agree and each is confident, and it carries no learned weights. Separately, when a single fused posterior is needed, for the closed-set prediction and for the probability-based novelty score, the per-class probability vectors of the individual classifiers are combined by one of two simple non-trainable rules, the arithmetic mean of the probabilities, called the mean rule, or their normalized element-wise product, called the product rule. Neither rule is a trained stacking model and neither uses tuned weights, which keeps the fusion transparent and avoids adding parameters that would themselves require validation.

\subsection{Hybrid Open-Set Novelty Detection}
\label{sec:hybrid}

The second stage examines the rejected samples to separate ambiguous known samples from genuinely novel odors, and it does so with two independent criteria whose agreement is required before a decision is taken.

The first criterion is geometric and uses the feature space learned by the network. The sample is mapped to its penultimate-layer embedding $z_{\text{test}}$, and a novelty score is computed as the squared Mahalanobis distance to the nearest known-class centroid $\mu_k$, as in Equation~\eqref{eq:novelty}. The covariance is pooled across the known classes rather than estimated separately for each one, since a single shared estimate is far better conditioned than several separate ones would be from only a few hundred training embeddings per class. It is obtained by centering every training embedding on its own class centroid and averaging the resulting outer products over the whole training set, as in Equation~\eqref{eq:covariance}, where $N$ is the number of training embeddings in that fold and $\mu_{y_i}$ the centroid of the class $y_i$ that training sample $i$ belongs to, after which a ridge term of $10^{-6}$ is added to its diagonal for numerical stability and the regularized matrix is inverted once per fold. If the resulting score exceeds an adaptive threshold $\tau$, the sample is flagged as a potential novelty. The threshold $\tau$ is set to the ninety-fifth percentile of this score computed on a held-out set of known-class samples, the same five percent operating point on the known classes used consistently throughout the paper, and it is calibrated using only those held-out known samples, before any withheld or unknown-class sample is scored or examined. The use of a covariance-normalized distance to class centroids in a learned embedding follows the prototype and distance-based detectors that underpin modern open-set recognition \citep{bendale_openmax, liang_odin, lee_mahalanobis}.

\begin{equation}
    \Sigma = \frac{1}{N}\sum_{i=1}^{N} \left(z_i - \mu_{y_i}\right)\left(z_i - \mu_{y_i}\right)^{\top}
    \label{eq:covariance}
\end{equation}

\begin{equation}
    \begin{aligned}
    S(x_{\text{test}}) &= \min_{k}\, \left(z_{\text{test}} - \mu_k\right)^{\top} \\
    &\quad {}\times \left(\Sigma + \epsilon I\right)^{-1} \left(z_{\text{test}} - \mu_k\right), \\
    &\quad \epsilon = 10^{-6}, \quad I \text{ the identity matrix}
    \end{aligned}
    \label{eq:novelty}
\end{equation}

The second criterion uses a lightweight disparity network. It operates on the 14 standardized sensor features, not on the five-dimensional DNN bottleneck. Its architecture is $14\to16\to8\to1$. Both hidden layers use rectified linear activations, and the output layer uses a sigmoid. The network therefore contains 385 trainable parameters: $14\times16+16$, $16\times8+8$, and $8\times1+1$. We train it with binary cross-entropy, Adam, a learning rate of $0.02$, a weight decay of $10^{-4}$, and 300 epochs. We construct all training inputs in the 14-dimensional standardized sensor feature space of that fold. We generate 80 points by interpolating between known-class centroids with a mixing ratio drawn from $[0.35,0.65]$. We generate another 200 points by extrapolating training feature vectors from their class centroids by a factor drawn from $[1.8,3.0]$. We regenerate these 280 boundary points for every fold and never use the withheld class to train the disparity network. We calibrate $\theta_D$ as the 95th percentile of its output on held-out known samples. Requiring the geometric and disparity criteria to agree reduces false alarms relative to either test alone.

\subsection{Active Adaptation and Incremental Class Creation}
\label{sec:classcreate}

The third stage turns a confirmed detection into a controlled update. A sample is treated as a novel odor only when both the geometric test and the lightweight verifier flag it, and this dual check keeps spurious alarms low. Once a sample is confirmed, the system enters an active-learning state rather than retraining immediately on potentially noisy data. The novel sample and subsequent similar samples are quarantined in a buffer, which prevents the system from reacting to transient sensor spikes or to single anomalies. In supervised deployments these high-priority samples can be shown to a human expert for labeling, which uses scarce expert effort efficiently. When enough quarantined samples have accumulated, a new class $n_K+1$ is provisionally created, and the ensemble is incrementally retrained to include it. For future deployment, we propose an additional cross-validation step after class creation to assess separation of the candidate class before releasing the updated models. We did not perform this additional step in the experiments reported here. Section~\ref{sec:adaptive_validation} instead evaluates a predefined sequence of training, streaming, updating, and testing, with a reserved segment for recognition of the new class. Repeating that sequence across five predefined folds is distinct from adding a validation step after each update. In practice the incremental training step is triggered only after several or a dozen samples of the candidate class have been collected, and once a small set is available the network can be extended to recognise the new class even from a few-shot presentation \citep{Wanyan_few, liu_deepsniffer, Snell_NeurIPS}. This cycle is intended to support adaptation with periodic expert verification. Its performance in a live deployment remains to be evaluated.

\subsection{Exploratory Clustering and Visualisation}

Clustering plays a fundamental role in the analysis of e-nose data, especially when labels are scarce or when unknown classes are being explored. Classical algorithms such as $k$-means \citep{kmeans_book}, fuzzy $c$-means \citep{cmeans_book}, and hierarchical clustering \citep{hier_clust} are widely used for their simplicity and interpretability, while more recent deep-clustering methods learn an embedding and cluster within it, which can reveal latent odor structure and help compensate for drift \citep{xie_dec}. In this work, $k$-means was used to group the stabilised sensor responses, and the cluster centroids were retained so that the distance between a new sample and the centroids of the established classes could be measured directly. To visualise the high-dimensional responses, principal component analysis projected the stabilised values onto two principal components \citep{pca_book}, which makes the separability of the odor classes easy to inspect.

\section{Results and Discussion}

\subsection{Experimental Protocol}

The acorn dataset contains 1,980 measurements, 330 from each of six classes. Each measurement is a 14-dimensional baseline normalized steady-state vector. We used leave-one-class-out evaluation, withholding each pathogen in turn as unknown. The other four pathogens, together with the medium, formed the five known classes. The raw file contains contiguous blocks of highly autocorrelated readings. We divided each known class into five contiguous segments and assigned whole segments to folds. This blocked split reduces leakage between adjacent readings. It cannot guarantee independence between cultures or acquisition sessions because the file lacks those identifiers. We fitted feature standardization only on each training fold and applied it to the corresponding test fold. We report closed-set accuracy and macro averaged F1 on held-out known samples. We report open-set AUROC, true-positive and false-positive rates, normalized accuracy, and open-set classification rate. We evaluated raw and temperature-scaled softmax, an energy score, OpenMax, and Mahalanobis distance in the five-unit bottleneck \citep{guo_calibration, liu_energy, bendale_openmax, lee_mahalanobis}.

\subsection{Exploratory Analysis and Qualitative Demonstration}

\begin{figure}[htbp]
    \centering
    \subfloat[]{\includegraphics[width=0.47\columnwidth]{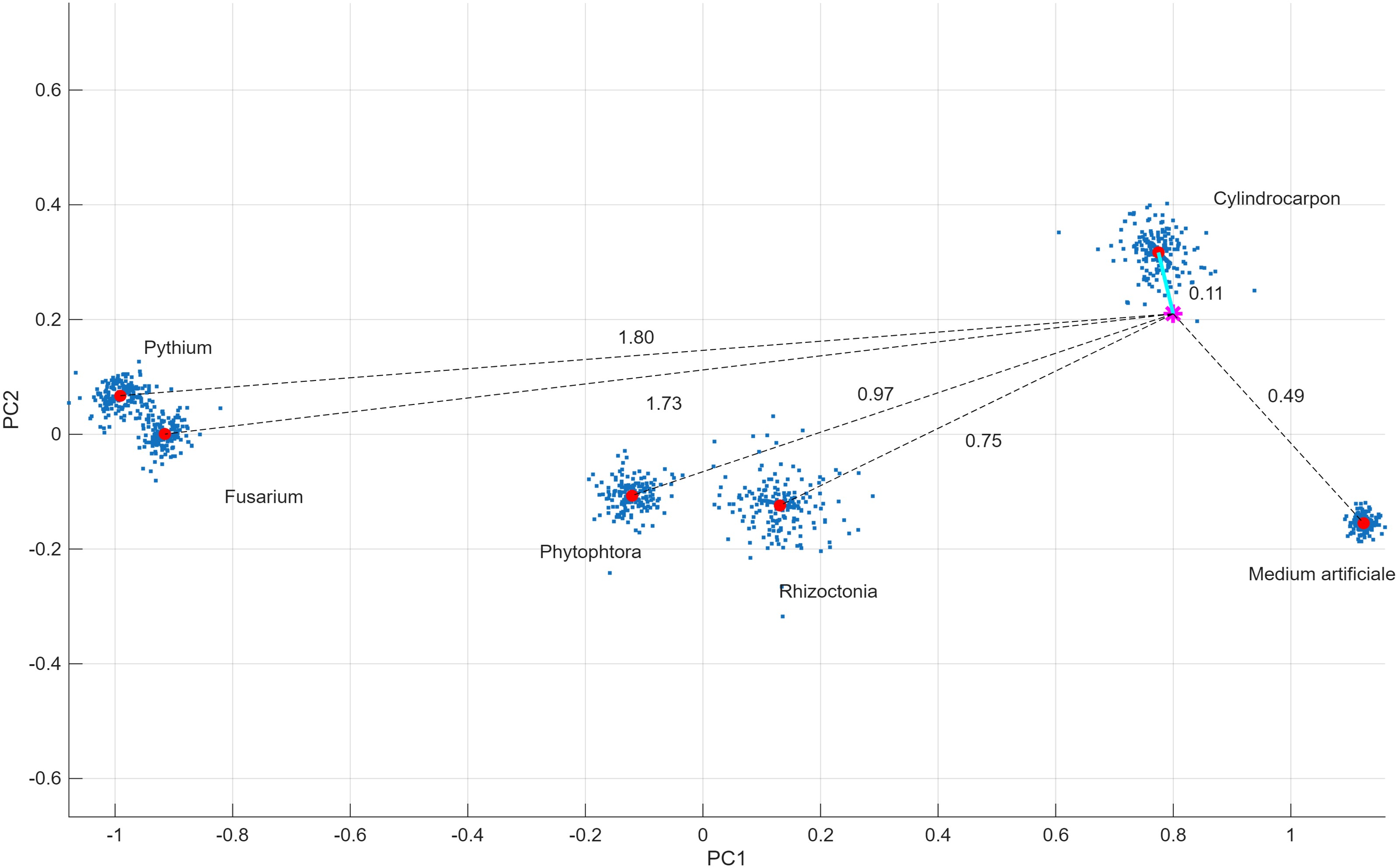}\label{fig:grzyby_pca2}}
    \hfill
    \subfloat[]{\includegraphics[width=0.47\columnwidth]{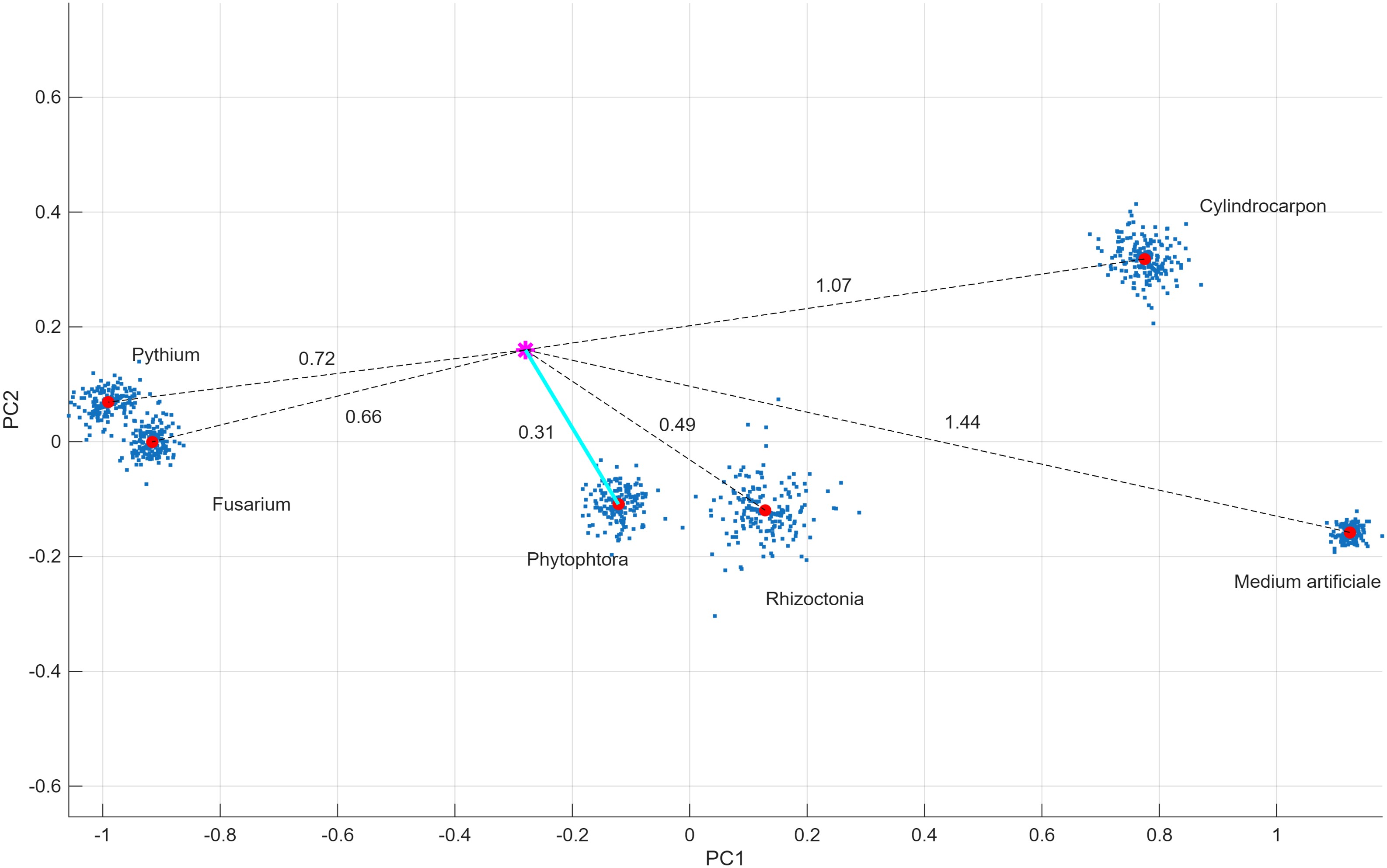}\label{fig:grzyby_pca1}}
    \caption{Qualitative principal-component illustration. The blue clouds and red centroids come from the main 1,980-sample acorn dataset. The magenta markers are additional measurements from a separate qualitative trial and were excluded from every quantitative evaluation. We selected the two examples deliberately to show a near-centroid case in panel (a) and a remote case in panel (b); they are not random or representative test samples.}
    \label{fig:grzyby_pca}
\end{figure}

Figures~\ref{fig:grzyby_pca2} and~\ref{fig:grzyby_pca1} illustrate the dimensionality reduction obtained with principal component analysis applied to the stabilised sensor values \citep{pca_book}. The two-component projection makes the separability of the odor classes visible. The clustering was performed with $k$-means \citep{kmeans_book}, which grouped the samples by their VOC profiles, and the cluster centroids are marked so that the distance between a new sample and the centroids of the established classes can be read directly from the figures.

Figure~\ref{fig:grzyby_pca2} shows the additional measurement near an existing centroid. Figure~\ref{fig:grzyby_pca1} shows the additional measurement outside the main data clusters. These panels illustrate how centroid distance behaves; they provide no quantitative evidence and do not contribute to any reported metric.

In the latter case, further analysis is warranted to decide whether the sample should be discarded as noise or used as the basis for a new odor category. That decision may call for additional sampling, expert validation, or retraining of the model to accommodate the expanded odor space. If the sample is confirmed to be a valid measurement that represents a new class, it is stored in the database, and once more samples of the same kind have been collected the network is updated to recognise the new class, which can be achieved even from a few-shot presentation \citep{Wanyan_few, Snell_NeurIPS}. This loop is intended to incorporate newly identified classes as samples accumulate. The present simulations evaluate individual updates and do not establish continuous improvement or resistance to drift and contamination during deployment. This update cycle is governed by fixed consensus and novelty thresholds rather than a learned decision policy, so it is described here as an incremental update mechanism rather than as reinforcement learning, a term that would require an explicit action space and reward signal that the framework does not define and that is not implemented or evaluated in this work. It is especially useful where olfactory profiles are non-stationary, as in agriculture, food safety, and clinical diagnostics.

\subsection{Quantitative Evaluation}

\begin{figure}[htbp]
    \centering
    \includegraphics[width=0.62\columnwidth]{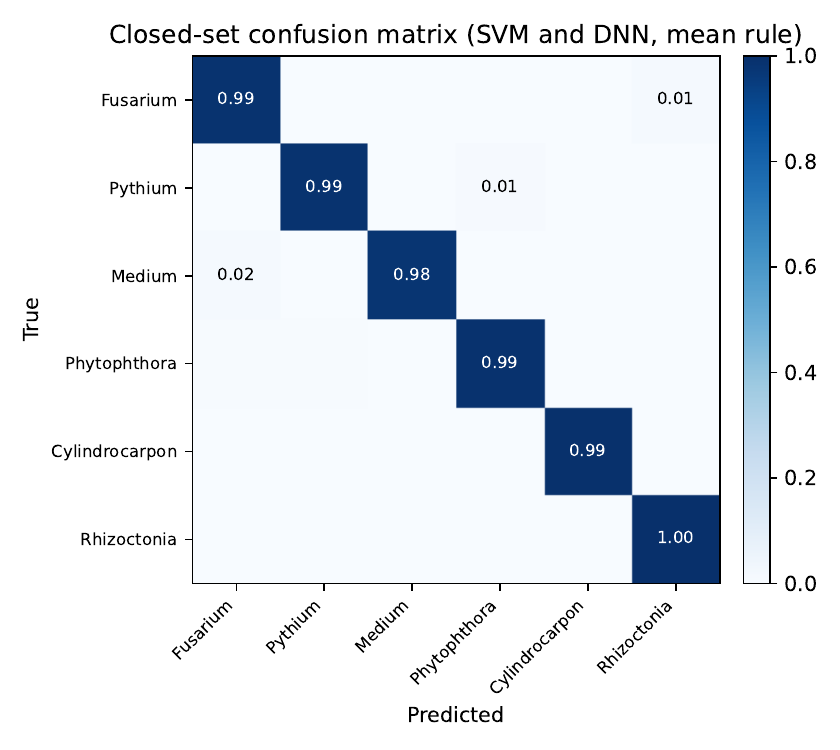}
    \caption{Row-normalized closed-set confusion matrix for the mean-rule fusion of the support vector machine and the network, aggregated over the leave-one-class-out folds. The known classes are separated almost perfectly.}
    \label{fig:confusion}
\end{figure}

On the known classes the task is close to saturated. Every classifier reaches between 0.986 and 0.996 in both overall accuracy and the macro-averaged F1 score, with the random forest highest, and the per-class precision and recall lie between 0.97 and 1.00 for all six classes. The only notable confusion is about two percent of medium samples assigned to \textit{Fusarium}, which is visible in the row-normalized confusion matrix of Figure~\ref{fig:confusion}. Because the known classes are so easily separated, closed-set accuracy does not distinguish the methods and is not where the contribution of the framework lies.

The per-class precision behind this confusion matrix ranges from 0.973 to 0.997 and recall from 0.979 to 0.996, with no class falling behind the others, which confirms that the closed-set difficulty is evenly low across the six classes rather than concentrated in one hard category, and the only confusion large enough to be visible in the figure is the two percent of medium samples the matrix already shows being assigned to \textit{Fusarium}.

The methods differ once the task becomes open-set, and Table~\ref{tab:results} collects the figures. Reading novelty from the network's softmax confidence gives an area under the ROC curve of 0.567, close to chance, because the network stays almost maximally confident on the withheld class just as it is on the known ones. Temperature scaling does not repair this and the energy score performs no better, which is consistent with the established observation that calibration alone is not out-of-distribution detection \citep{guo_calibration, liu_energy, hendrycks_msp}. The support vector machine and the random forest, whose confidences fall on unfamiliar inputs, give strong novelty scores with areas under the curve of 0.958 and 0.939. The established open-set baseline, OpenMax \citep{bendale_openmax}, which recalibrates the network logits by their distance to the class mean activation vectors and fits a Weibull tail to the training distances so that outliers are flagged, reaches an AUROC of 0.928 in this experiment, above raw softmax but below the observed values for SVM, random forest, mean fusion, and bottleneck distance. The principal result is that the same network attains the highest area under the curve of all methods, 0.971 and a true-positive rate of 0.873 at a five percent false-alarm rate, once its novelty is read from a Mahalanobis distance in the five-unit bottleneck rather than from its softmax \citep{lee_mahalanobis}. These observed values motivate the geometric criterion used in the framework. They do not establish its superiority over the alternatives, because the tested comparisons are not statistically significant after Holm correction.

\begin{table}[t]
\caption{Closed-set and open-set performance under the leave-one-class-out protocol, averaged over the five withheld pathogens and five blocked cross-validation folds. Accuracy and macro-F1 are measured on the held-out samples of the known classes and are shared by the four network rows because they use the same classifier. The novelty operating point is fixed at a five percent known false-positive rate, at which the true-positive rate (TPR) and the realized false-positive rate (FPR) are read. NA is the normalized accuracy with equal weight on knowns and unknowns. Higher is better.}
\label{tab:results}
\begingroup\small\setlength{\tabcolsep}{4pt}
\begin{tabularx}{\textwidth}{@{}lcccCCC@{}}
\toprule
\textbf{Method} & \textbf{Acc.} & \textbf{Macro-F1} & \textbf{AUROC} & \textbf{TPR} & \textbf{FPR} & \textbf{NA} \\
\midrule
DNN, softmax & 0.986 & 0.986 & 0.567 & 0.328 & 0.052 & 0.633 \\
DNN, temperature & 0.986 & 0.986 & 0.529 & 0.071 & 0.011 & 0.525 \\
DNN, energy & 0.986 & 0.986 & 0.380 & 0.172 & 0.052 & 0.554 \\
DNN, Mahalanobis & 0.986 & 0.986 & 0.971 & 0.873 & 0.052 & 0.906 \\
OpenMax \citep{bendale_openmax} & 0.986 & 0.986 & 0.928 & 0.800 & 0.052 & 0.868 \\
SVM (RBF) & 0.991 & 0.991 & 0.958 & 0.835 & 0.052 & 0.889 \\
Random forest & 0.996 & 0.996 & 0.939 & 0.817 & 0.041 & 0.887 \\
$k$-nearest neighbors & 0.993 & 0.993 & 0.643 & 0.306 & 0.014 & 0.644 \\
SVM+DNN, mean rule & 0.988 & 0.988 & 0.952 & 0.815 & 0.052 & 0.879 \\
SVM+DNN, product rule & 0.989 & 0.989 & 0.763 & 0.488 & 0.052 & 0.715 \\
\bottomrule
\end{tabularx}
\endgroup
\end{table}

\begin{figure}[htbp]
    \centering
    \includegraphics[width=0.8\columnwidth]{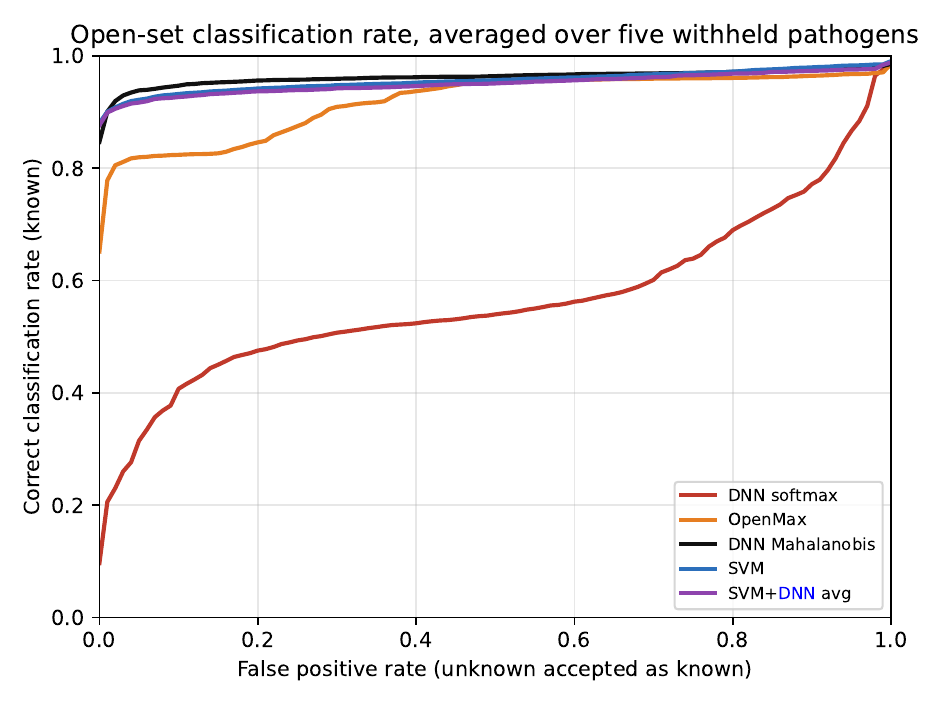}
    \caption{Open-set classification rate, the correct classification rate on the known classes against the rate at which unknown samples are wrongly accepted as known, averaged over the five withheld pathogens. The softmax detector lies far below the others, OpenMax recovers much of the gap, and the bottleneck Mahalanobis distance matches the support vector machine and the mean-rule fusion at the top.}
    \label{fig:oscr}
\end{figure}

Figure~\ref{fig:oscr} shows the same ordering through the open-set classification rate curve, on which the softmax detector trails far below the distance-based and margin-based detectors across the whole range of operating points, while the bottleneck distance, the support vector machine, the random forest, and the mean-rule fusion are almost indistinguishable near the top. The per-pathogen breakdown in Table~\ref{tab:perclass} shows the same pattern per class, since the softmax score falls below chance for \textit{Cylindrocarpon} and \textit{Rhizoctonia}, meaning the network is more confident on those unknown samples than on the known classes, whereas the bottleneck distance stays above 0.93 for every withheld pathogen. The two fusions of the support vector machine and the network behave as the individual scores predict. Averaging the probabilities preserves almost all of the support vector machine's novelty strength and reaches an area under the curve of 0.952, while the product rule is dragged down to 0.763 because the overconfident softmax dominates the product, so the mean rule has the higher observed AUROC in this experiment.

The spread across the five withheld pathogens shows that the leading detectors are not only the strongest but also the steadiest, which matters as much as the mean for a method that must work on a pathogen it has never seen. The bottleneck distance, the support vector machine, and the mean-rule fusion score 0.971 plus or minus 0.026, 0.958 plus or minus 0.025, and 0.952 plus or minus 0.026 respectively. The random forest at 0.939 plus or minus 0.063 and OpenMax at 0.928 plus or minus 0.069 are close behind but more variable from one pathogen to the next, with OpenMax dropping to 0.794 on its weakest case, and the $k$-nearest-neighbor baseline is markedly weaker at 0.643, which shows that a generic distance to neighbors is not enough and that the structured bottleneck representation is what makes the geometric criterion work. The confidence-based scores are weaker and less consistent, since the softmax at 0.567 plus or minus 0.335 and the energy score at 0.380 plus or minus 0.325 are not merely poor on average but wildly inconsistent, collapsing below chance for some pathogens, which is precisely the unpredictability a deployable detector must avoid.

The 25 pathogen-by-fold combinations are dependent on each other because they share training data. We treat the withheld pathogen as the independent unit. This gives 5 pathogen-level means per method. The 95\% cluster-bootstrap intervals are $[0.947,0.993]$ for bottleneck distance, $[0.936,0.980]$ for SVM, $[0.930,0.975]$ for mean-rule fusion, $[0.876,0.985]$ for random forest, and $[0.857,0.975]$ for OpenMax. These intervals overlap. Holm-corrected paired Wilcoxon tests find no statistically significant pairwise difference among the tested methods. With only 5 independent units, the tests have little inferential resolution. We therefore treat them as exploratory and emphasize the descriptive AUROCs, intervals, and directions across pathogens. Bottleneck distance ranks above softmax for all five pathogens ($r=1.00$). Mean-rule fusion ranks above product-rule fusion for all five ($r=1.00$). SVM ranks above $k$-nearest neighbors for all five ($r=1.00$). In contrast, bottleneck distance exceeds SVM in only three pathogens ($r=0.60$) and OpenMax in three ($r=0.47$). We therefore interpret bottleneck distance, SVM, random forest, mean-rule fusion, and OpenMax as a leading group rather than a strict ranking.

\begin{table}[t]
\caption{Novelty AUROC when each pathogen in turn is treated as the unknown class. Values below 0.5 mean the detector is more confident on the unknown samples than on the known ones. The Mahalanobis score on the network bottleneck is strong for every pathogen, whereas the softmax and energy scores collapse for several of them.}
\label{tab:perclass}
{\footnotesize\centering\setlength{\tabcolsep}{4pt}
\begin{tabular}{@{}lccccc@{}}
\toprule
\textbf{Method} & \textit{Fusarium} & \textit{Pythium} & \textit{Phytophthora} & \textit{Cylindrocarpon} & \textit{Rhizoctonia} \\
\midrule
DNN, softmax & 0.926 & 0.760 & 0.822 & 0.107 & 0.222 \\
DNN, temperature & 0.616 & 0.515 & 0.526 & 0.489 & 0.498 \\
DNN, energy & 0.962 & 0.163 & 0.344 & 0.006 & 0.427 \\
DNN, Mahalanobis & 0.984 & 0.953 & 0.931 & 0.998 & 0.992 \\
OpenMax \citep{bendale_openmax} & 0.940 & 0.995 & 0.952 & 0.794 & 0.960 \\
SVM (RBF) & 0.991 & 0.928 & 0.937 & 0.953 & 0.982 \\
Random forest & 0.987 & 0.822 & 0.971 & 0.924 & 0.990 \\
$k$-nearest neighbors & 0.705 & 0.513 & 0.534 & 0.808 & 0.654 \\
SVM+DNN, mean rule & 0.990 & 0.936 & 0.928 & 0.929 & 0.975 \\
SVM+DNN, product rule & 0.963 & 0.929 & 0.891 & 0.388 & 0.643 \\
\bottomrule
\end{tabular}
}
\end{table}

\subsection{Robustness to Sensor Drift on a Public Benchmark}
\label{sec:external}

The acorn study describes performance on the available biological dataset. Independence between cultures or acquisition sessions could not be established. We therefore also evaluated novelty detection on an external dataset recorded with different hardware across temporal batches. The Gas Sensor Array Drift Dataset of Vergara and colleagues \citep{vergara_drift, vergara_dataset} serves as that stress test. It contains 13{,}910 measurements from an array of sixteen metal-oxide sensors of the same Figaro family used here, exposed to six analytes over thirty-six months, and the recordings are grouped into ten temporal batches that carry the drift explicitly, which also makes the cross-validation strictly session-grouped. Each batch is held out in turn while the models train on the other nine, and within each split one analyte is withheld as the unknown class, giving up to sixty combinations of held-out batch and withheld analyte in principle. Three of these sixty do not exist in the data and are excluded, all three involving toluene, whose measurements are entirely absent from batches three, four, and five of the original dataset, leaving zero samples of that analyte to serve as the unknown class in those three batches. No other combination was excluded, and every other analyte is present in every one of the ten batches, so the remaining fifty-seven combinations are all genuinely valid. The network, the support vector machine, the random forest, the $k$-nearest-neighbor baseline, and the mean-rule fusion are scored exactly as before, with the network taking the fully connected form of Section~\ref{sec:nn} for the 128-dimensional descriptors. As on the acorn data, the novelty threshold for every method is calibrated independently within each training partition, from that partition's own held-out known-class samples only, before the withheld analyte's samples are examined, so no combination's threshold depends on data outside its own split.

\begin{figure}[htbp]
    \centering
    \includegraphics[width=0.95\columnwidth]{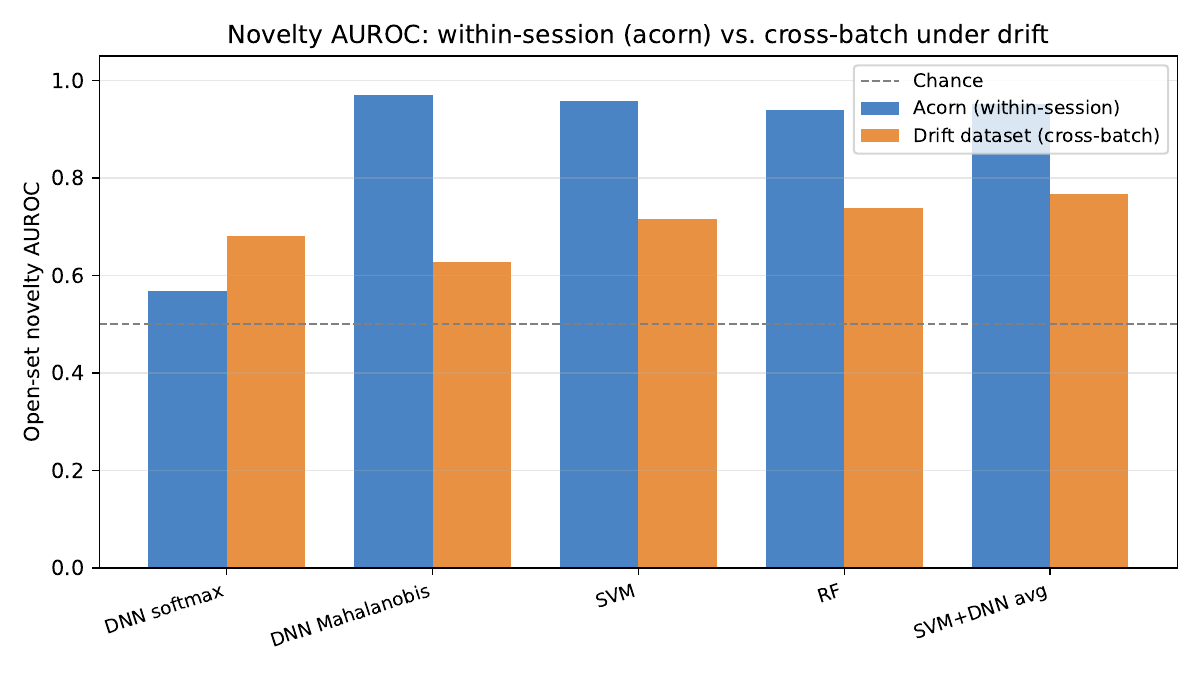}
    \caption{Open-set novelty AUROC in the acorn study and the cross-batch drift benchmark. The legend label ``SVM+DNN avg'' denotes the arithmetic mean-rule fusion of the SVM and DNN probabilities. Bottleneck distance has the highest observed mean AUROC on the acorn data, while mean fusion has the highest observed value on the drift benchmark. The comparisons after Holm correction are not statistically significant.}
    \label{fig:drift_cmp}
\end{figure}

Two effects appear. Closed-set accuracy falls from near unity to between 0.85 and 0.89, which is expected once training on nine batches and testing on a tenth exposes the classifier to drift directly. The open-set ranking reverses, as shown in Figure~\ref{fig:drift_cmp}. Bottleneck Mahalanobis distance had the highest observed mean AUROC on the acorn data, 0.971, and reached 0.628 on this benchmark. The corresponding values were 0.680 for raw softmax, 0.715 for SVM, 0.739 for random forest, and 0.768 for mean fusion. These values describe the present experiments and do not establish a statistically significant ranking. As with the acorn study, the fifty-seven gas-by-batch combinations are not independent, since different withheld gases evaluated at the same held-out batch share most of their known-class training data, so the withheld gas, not the combination, is treated as the independent unit here as well, giving six values per method, five gases with ten valid batches each and toluene with seven. The differences are directionally consistent rather than an averaging artefact, since the ninety-five percent confidence interval, obtained the same way by a cluster bootstrap over the six withheld gases, is $[0.586, 0.879]$ for the fusion against $[0.520, 0.746]$ for the bottleneck distance. These intervals overlap enough that a paired Wilcoxon signed-rank test on the six gas-level means, with a Holm-Bonferroni correction applied across the four comparisons reported here, does not reach significance for any of them, the fusion against the bottleneck distance (adjusted $p = 0.88$), the fusion against the softmax (adjusted $p = 0.88$), the random forest against the bottleneck distance (adjusted $p = 0.88$), and the support vector machine against the bottleneck distance (adjusted $p = 0.88$). The rank-biserial effect sizes still show a consistent direction, the fusion ranks above the bottleneck distance in five of the six withheld gases ($r = 0.62$) and above the softmax in five of six ($r = 0.62$), and the random forest and the support vector machine each rank above the bottleneck distance in five of six ($r = 0.43$ for both). The reversal relative to the acorn study is accordingly a consistent direction across most withheld gases rather than a formally significant effect at this cluster count, which is the honest limit of what six independent gases can establish.

Because several withheld-analyte combinations share the same test batch, the batch is a second, distinct source of dependence alongside the withheld gas, so it is worth checking directly rather than assumed away. Averaging within each of the ten batches instead of within each of the six gases shows a real but smaller effect, the mean-rule fusion's accuracy by batch ranges from 0.639 to 0.886, a standard deviation of 0.067 across batches, against 0.358 to 0.900 and a standard deviation of 0.189 across gases, so the withheld analyte is the larger source of variation between the two. A cluster bootstrap that resamples over the ten batches instead of the six gases confirms this rather than merely suggesting it, its interval is narrower than the gas-clustered interval already reported for every one of the five methods compared, so the intervals in this paper, which are built by resampling the larger and more conservative of the two dependence structures, are not understating the uncertainty that batch-level dependence would add. A possible explanation is that drift changes the relation between test samples and the centroids estimated during training. The present comparison does not isolate this mechanism. Bottleneck distance had the highest observed mean AUROC on the acorn data, whereas mean fusion had the highest observed value on the drift benchmark. These descriptive findings are specific to the datasets and protocols evaluated here. With five pathogen units and six gas units, and no significant comparisons after Holm correction, they do not establish a generally preferred detector.

\subsection{Validation of the Consensus, Novelty, and Class-Creation Cycle}
\label{sec:adaptive_validation}

The consensus gate of Section~\ref{sec:consensus}, the hybrid novelty test of Section~\ref{sec:hybrid}, and the class-creation cycle of Section~\ref{sec:classcreate} are evaluated here as a connected sequence rather than as isolated components, using the acorn data under the same blocked, five-fold structure as the rest of the study. Three questions are addressed in turn, what fraction of unlabeled samples the consensus gate accepts and how accurate the resulting pseudo-labels are, whether the disparity network contributes anything beyond the geometric criterion it is paired with, and whether a newly created class is recognised correctly while accuracy on the established classes is retained.

For the first question, each fold sets aside one contiguous segment per class as a clean test set untouched by the loop, keeps a small labeled seed from the remaining segments, and treats the rest as an unlabeled pool whose true labels are hidden from the algorithm and used only afterwards to score what the gate accepted. With a seed of fifteen percent of the available training data per class, the gate accepts a large majority of the pool and every accepted pseudo-label matches its hidden label, but closed-set accuracy is already close to its ceiling before any pseudo-labeled sample is added, so self-training changes almost nothing, as Table~\ref{tab:selftrain} shows. Reducing the seed to about three percent of the available data per class produces a markedly different and less favourable picture. At the confidence threshold of ninety percent proposed in Section~\ref{sec:consensus}, the gate accepts nothing at all in any of the five folds, because with so few training samples the calibrated probability of the support vector machine rarely exceeds ninety percent for all three models at once. Lowering the threshold to eighty percent, the other end of the range proposed there, does not repair this so much as make the gate unpredictable, accepting nothing in two folds, almost the entire pool in one, and a moderate fraction in the remaining two. Pseudo-label precision remains at or near unity whenever the gate accepts anything, so the mechanism is trustworthy when it fires, but whether it fires at all depends strongly on how much labeled data is available at the outset and on the exact threshold chosen, a sensitivity the original description does not discuss.

\begin{table}[t]
\caption{Consensus-gated self-training under three seeding and threshold regimes, averaged over five blocked folds. Precision is computed only over samples the gate actually accepted, and is undefined when nothing is accepted.}
\label{tab:selftrain}
\begin{tabularx}{\textwidth}{lSSSS}
\toprule
\textbf{Regime} & \textbf{Acceptance rate} & \textbf{Pseudo-label precision} & \textbf{Accuracy before} & \textbf{Accuracy after} \\
\midrule
Seed 15\%, threshold 90\% & 0.857 & 1.000 & 0.989 & 0.989 \\
Seed 3\%, threshold 90\% & 0.000 & undefined & 0.978 & 0.978 \\
Seed 3\%, threshold 80\% & 0.282 & 1.000 & 0.978 & 0.977 \\
\bottomrule
\end{tabularx}
\end{table}

For the second question, we calibrate the Mahalanobis and disparity criteria separately to a 5\% false-positive rate on held-out known samples. We then compare each criterion against the rule requiring both to agree. The disparity network has 385 parameters and uses the fourteen standardized sensor features described in Section~\ref{sec:hybrid}. Requiring agreement reduces the false-positive rate from about 5\% to 0.8\%. The true-positive rate also falls from 87.3\% for the Mahalanobis distance to 66.8\% for the hybrid rule. The hybrid rule increases specificity at the cost of sensitivity.

\begin{table}[t]
\caption{Contribution of the disparity network to the hybrid novelty test, averaged over the same twenty-five combinations of withheld pathogen and fold as Table~\ref{tab:results}. Each criterion is calibrated separately to a five percent false-positive rate on held-out known samples.}
\label{tab:disparity}
\begin{tabularx}{\linewidth}{lSS}
\toprule
\textbf{Criterion} & \textbf{False-positive rate} & \textbf{True-positive rate} \\
\midrule
Mahalanobis distance alone & 0.052 & 0.873 \\
Disparity network alone & 0.052 & 0.675 \\
Hybrid, both required & 0.008 & 0.668 \\
\bottomrule
\end{tabularx}
\end{table}

\begingroup
For the third question, we fixed the train, stream, update, and test partitions before processing any stream sample. We did not run an additional cross-validation step after class creation; the reserved segment provided the evaluation of the new class after each completed update. For each known class, four contiguous segments fitted the initial ensemble. The remaining known class segment calibrated the novelty thresholds and was then interleaved with the deployment stream. For the withheld pathogen, four segments entered that stream, and the fifth segment was reserved for post-update testing. Every stream sample flagged by both novelty criteria entered the quarantine buffer, including any false alarms. When the buffer reached 20 samples, we labeled that buffer as the candidate new class. We retrained the ensemble on the original known-class training data plus those 20 buffered samples. The reserved withheld pathogen segment and later stream samples were not included in retraining. New class recognition was measured on the reserved withheld pathogen segment, which was excluded from calibration, streaming, and updating. Known class retention was measured on the known class segment that had also been used for calibration and in the stream. It is therefore a descriptive diagnostic, not an independent post-update test. We make this reuse explicit rather than claiming an independent known class retention evaluation.

The buffer filled in 19 of 25 pathogen-by-fold simulations, a class creation trigger rate of 76\%. We calculated buffer purity, retention, and new-class recognition only for those 19 completed updates. Their mean new class recognition on the reserved withheld-pathogen segment was 98.2\%. The descriptive known class retention diagnostic changed from 98.7\% before updating to 99.3\% afterward. The other six simulations did not create a class and do not contribute to the 98.2\% value. Four failures occurred in fold 5 and two in fold 4. Fold 5 also had the lowest pre-update closed-set accuracy. Because the acquisition order cannot be linked to cultures or sessions, we do not assign a biological or related to acquisition cause to this concentration.
\endgroup

\begin{table}[t]
\caption{Quarantine and class-creation results across 25 pathogen-by-fold simulations. The trigger rate is $19/25=76\%$. Buffer purity, retention, and new-class recognition describe only the 19 simulations that completed class creation. New-class recognition uses a withheld-pathogen segment excluded from the stream and updates. The known-class retention fold was reused for threshold calibration and stream simulation, so retention is descriptive rather than independent.}
\label{tab:classcreate}
\begin{tabularx}{\linewidth}{lS}
\toprule
\textbf{Quantity} & \textbf{Value} \\
\midrule
Combinations that filled the buffer & 19 / 25 \\
Median stream samples needed & 54 \\
Buffer purity & 0.971 \\
Retention accuracy, before the update & 0.987 \\
Retention accuracy, after the update & 0.993 \\
New-class recognition accuracy & 0.982 \\
\bottomrule
\end{tabularx}
\end{table}

\begin{table}[t]
\caption{Outcome of the class-creation cycle by fold. The fold with the lowest closed-set accuracy before adaptation is also the fold where the cycle most often fails to fill the buffer within the available stream.}
\label{tab:classcreate_fold}
\begin{tabularx}{\linewidth}{lSS}
\toprule
\textbf{Fold} & \textbf{Combinations filled} & \textbf{Closed-set accuracy before adaptation} \\
\midrule
1 & 5 / 5 & 1.000 \\
2 & 5 / 5 & 1.000 \\
3 & 5 / 5 & 1.000 \\
4 & 3 / 5 & 0.999 \\
5 & 1 / 5 & 0.936 \\
\bottomrule
\end{tabularx}
\end{table}

These results separate two outcomes. The class creation trigger succeeds in 76\% of simulations. Conditional on a successful trigger, the updated model recognizes the new class with 98.2\% accuracy on a segment excluded from the stream and update. The known class retention diagnostic remains high, but it is not an independent post-update estimate because its fold was reused during calibration and streaming. The failures concentrate in the empirically weakest fold, but the available metadata do not support a causal explanation.

\subsection{Computational Cost and Error Accumulation Across Adaptation Rounds}
\label{sec:cost_and_error}

Two further properties of the framework are measured directly here rather than assumed, since neither was checked in the experiments above: the computational cost of producing a prediction once training is complete, and whether pseudo-label errors compound as the self-training loop runs across successive rounds.

Table~\ref{tab:latency} reports the wall-clock time to classify and score one new sample with each already-trained component, measured on a single core of a generic Intel Xeon server processor at 2.10\,GHz, the platform this study was carried out on rather than a purpose-built embedded target. The deep network, the support vector machine, the Mahalanobis novelty score, and the disparity network are all sub-millisecond per sample. The random forest is the dominant cost, at roughly fifteen milliseconds per sample, because scikit-learn's implementation is optimized for scoring many samples in a batch rather than one at a time, and calling it on a single sample repeatedly incurs per-call overhead that a batched or embedded implementation would not. Even so, the complete pipeline, all three classifiers and the hybrid novelty test together, classifies a new sample in about sixteen milliseconds, several orders of magnitude faster than the many-second sensor exposure and stabilisation time a single measurement itself requires, as shown by the response curve of Figure~\ref{fig:signal_sensors}. Computation is accordingly not the bottleneck to deployment on this hardware, though this remains a desktop-class reference point rather than a measurement on power- or memory-constrained embedded hardware, which would need its own separate characterisation.

\begin{table}[t]
\caption{Wall-clock inference latency per sample for each component of the framework, measured on a single core of a generic Intel Xeon server processor at 2.10\,GHz after training is complete. The random forest dominates the total because of per-call overhead in scoring one sample at a time rather than a batch.}
\label{tab:latency}
\begin{tabularx}{\textwidth}{lLL}
\toprule
\textbf{Component} & \textbf{ms / sample} & \textbf{samples / s} \\
\midrule
DNN, softmax forward pass & 0.045 & 22{,}200 \\
SVM, RBF kernel & 0.070 & 14{,}300 \\
Mahalanobis novelty score (incl.\ DNN feature pass) & 0.067 & 14{,}800 \\
Disparity network forward pass & 0.010 & 96{,}700 \\
Random forest & 15.407 & 65 \\
\midrule
Full pipeline (all three classifiers plus hybrid novelty test) & 16.058 & 62 \\
\bottomrule
\end{tabularx}
\end{table}

On error accumulation, every self-training round reported in Table~\ref{tab:selftrain}, across all five folds and all three seeding and threshold regimes, was checked individually against the hidden true labels rather than only the aggregate figures already given. Across the fifty-one rounds in which the consensus gate accepted at least one sample, it accepted seven thousand nine hundred and sixty pseudo-labeled samples in total, of which exactly one was incorrect, an overall precision of 0.9999 across every round examined, and the single error occurred at the final round of one fold under the smaller seed and the more permissive threshold, the regime already identified as the least reliable. Pseudo-label errors are accordingly not compounding across successive rounds in these experiments, the gate's precision holds essentially unchanged from the first round to the last, though this conclusion is limited to the up to eight rounds and the pool sizes studied here, and a longer-running deployment with repeated class-creation events over more rounds than tested would still warrant its own monitoring.
\subsection{Advantages of the Proposed Framework}

The framework combines classification, novelty detection, and provisional class creation in a single procedure. The reported simulations demonstrate these operations on the available data. They do not establish improved generalization to independently prepared cultures or acquisition sessions, because those groups could not be reconstructed. The consensus gate achieved high precision among accepted pseudo-labels, while the hybrid novelty rule reduced false alarms at the cost of sensitivity. Learning from unlabeled samples may reduce manual labeling needs, but the practical benefit requires evaluation in deployment. Section~\ref{sec:adaptive_validation} provides direct evidence for each stage. The consensus gate maintains high pseudo-label precision when it accepts samples. The hybrid rule reduces false alarms below 1\%. Class creation triggers in 19 of 25 simulations. In those 19 completed updates, new class recognition reaches 98.2\% on a withheld pathogen segment excluded from streaming and updating. The known class retention result is descriptive because its fold was reused for calibration and streaming. These findings come from offline data, not a live deployment. Section~\ref{sec:cost_and_error} reports inference latency on the study hardware, but embedded deployment remains untested. These properties are valuable in agriculture, food safety, and environmental monitoring.

\subsection{Limitations}
\label{sec:limitations}

Several limitations should temper these claims and guide the next steps. The performance of the system depends on the quality of the initial labeled dataset, so a weak or biased starting set will limit later accuracy. Retraining several classifiers each time a class is added carries a computational cost, which matters for embedded or low-power deployment. The confidence and novelty thresholds must be tuned with care, since they govern the trade-off between false alarms and missed detections, and a single fixed value rarely suits every operating condition. A further caution concerns the data itself, and it can now be stated precisely rather than inferred. Each pathogen and the medium control were prepared as five independent cultures across five consecutive days, with three jars exposed within every culture, so every class rests on fifteen independently prepared jars rather than one, a real though modest amount of biological replication. This does not resolve the central concern, however, because the order in which the three hundred and thirty measurements of each class were written to the shared data file cannot be confirmed to follow the acquisition order. It is therefore not possible to identify which rows came from which culture or which jar, and a direct check for day-to-day structure within each class's readings found no signal distinguishing one contiguous block of the file from another, consistent with either an unordered file or with preprocessing that removed such structure before the file was compiled. Because the infected acorns used for this study are no longer available, this cannot be resolved by returning to the same material, and validation that holds out an entire culture rather than a contiguous file position is not possible on this dataset as it stands. The blocked cross-validation used throughout this paper therefore guards only against leakage between adjacent, highly autocorrelated readings drawn from the same brief exposure, not against leakage between readings that happen to share a culture or a jar, and the closed-set and open-set figures reported on the acorn data should be read accordingly, as evidence that the framework separates and detects departures among five known cultures per class, not as an estimate of its accuracy on an independently cultured new sample. The drift robustness analysis of Section~\ref{sec:external} is accordingly the strongest evidence in this paper for behavior beyond the specific material at hand, since its validation is genuinely grouped by acquisition batch on an independent public benchmark, and the acorn figures should be read together with that benchmark rather than on their own. A study designed from the outset to record culture and session identifiers alongside each measurement is the appropriate way to close this gap, and is a necessary precondition for the next study on this material rather than an optional refinement. Finally, the present study rests on a single biological material, and a comparison on the identical preprocessing used by Qu and colleagues \citep{qu_openset} remains a useful further step.

The adaptive validation adds several limitations. The consensus gate's acceptance rate depends on the labeled seed and the heuristic threshold. The hybrid rule reduces false alarms but also misses more novel samples. Class creation triggers in 76\% of simulations, while the 98.2\% recognition rate applies only after a successful trigger. The new-class test segment is excluded from streaming and updating, but the known-class retention fold is reused for threshold calibration and stream simulation. Retention is therefore descriptive rather than an independent post-update estimate. Failures concentrate in the weakest fold. We cannot connect file order to cultures or sessions, so we do not propose a biological or acquisition-related explanation. The validation uses an offline simulated stream. It does not establish performance in a live or embedded deployment.

\section{Summary and Conclusions}

This article presented a semi-supervised adaptive framework for electronic-nose systems that assigns odor samples to known categories and creates new categories when a sample departs from every known profile. The method couples a strict three-model consensus rule to a self-training loop, so that probabilistic, geometric, and statistical evidence must agree before an unlabeled sample is trusted, and it pairs this gate with a dual-criterion open-set detector and an active quarantine-and-adapt cycle that creates a new class only after verification. The aim throughout was to keep the strong closed-set accuracy of an ensemble while adding an explicit ability to recognise and learn the unknown, which is the central weakness of conventional supervised e-nose models.

We first studied the framework on volatiles emitted by root pathogens cultured on germinating \textit{Quercus robur} acorns. Leave-one-class-out evaluation used contiguous blocked splits to reduce leakage between adjacent autocorrelated readings. It could not establish culture level independence. Known class accuracy and macro-averaged F1 ranged from 0.99 to 1.00. For withheld classes, softmax confidence reached an AUROC of 0.567, OpenMax reached 0.928, SVM reached 0.958, and Mahalanobis distance in the five unit bottleneck reached 0.971. With five independent pathogen level units, these differences remain descriptive rather than confirmatory. The observed AUROC for feature space novelty was high on this dataset. This result does not establish performance on independent cultures or acquisition sessions.

On the public Gas Sensor Array Drift Dataset, we held out complete temporal batches. Closed-set accuracy ranged from 0.85 to 0.89. Mean fusion had the highest observed mean novelty AUROC, 0.768, compared with 0.628 for bottleneck distance. These descriptive results suggest that the relative performance of the detectors depends on the dataset and evaluation conditions. The comparisons after Holm correction were not statistically significant, so they do not establish a generally superior method. Evaluation on additional independent cultures and acquisition sessions is needed before making broader claims about generalization.

We then evaluated the adaptive stages as a connected sequence. The consensus gate maintained high pseudo-label precision when it accepted samples, but its acceptance rate depended on seed size and threshold. Combining geometric and disparity criteria reduced false alarms below 1\% and reduced sensitivity. Class creation triggered in 19 of 25 simulations. The 19 completed updates recognized the new class with 98.2\% accuracy on a segment excluded from streaming and updating. Known-class retention remained high, but the diagnostic reused the calibration and stream fold, so it is not independent. Failures concentrated in the empirically weakest fold, but the data do not support a causal explanation.

Several extensions follow naturally from the framework. The active class-creation cycle already supplies a hook for online recalibration of the class centroids, which is a promising route to carrying the feature-space advantage into changing conditions, and richer feature extractors for the raw transients together with optimization for low-power edge deployment would move the system toward autonomous, self-updating olfactory sensing in the field.






\bibliographystyle{unsrtnat}
\bibliography{bibi2}

@article{Wilson_enos,
  author    = {Wilson, Alphus D. and Baietto, Manuela},
  title     = {Applications and Advances in Electronic-Nose Technologies},
  journal   = {Sensors},
  volume    = {9},
  number    = {7},
  pages     = {5099--5148},
  year      = {2009},
  doi       = {10.3390/s90705099},
  url       = {https://doi.org/10.3390/s90705099}
}

@article{wilson2011applications,
  title     = {Advances in Electronic-Nose Technologies Developed for Biomedical Applications},
  author    = {Wilson, Alphus D. and Baietto, Manuela},
  journal   = {Sensors},
  volume    = {11},
  number    = {1},
  pages     = {1105--1176},
  year      = {2011},
  doi       = {10.3390/s110101105},
  url       = {https://doi.org/10.3390/s110101105}
}

@article{Ye_review,
  author    = {Ye, Zhenyi and Liu, Yuan and Li, Qiliang},
  title     = {Recent Progress in Smart Electronic Nose Technologies Enabled with Machine Learning Methods},
  journal   = {Sensors},
  volume    = {21},
  number    = {22},
  pages     = {7620},
  year      = {2021},
  doi       = {10.3390/s21227620},
  url       = {https://doi.org/10.3390/s21227620}
}

@article{Zhai_review,
  author    = {Zhai, Zhenyu and Liu, Yaqian and Li, Congju and Wang, Defa and Wu, Hai},
  title     = {Electronic Noses: From Gas-Sensitive Components and Practical Applications to Data Processing},
  journal   = {Sensors},
  volume    = {24},
  number    = {15},
  pages     = {4806},
  year      = {2024},
  doi       = {10.3390/s24154806},
  url       = {https://doi.org/10.3390/s24154806}
}

@article{Rock_enos,
  author    = {R{\"o}ck, Frank and Barsan, Nicolae and Weimar, Udo},
  title     = {Electronic Nose: Current Status and Future Trends},
  journal   = {Chemical Reviews},
  volume    = {108},
  number    = {2},
  pages     = {705--725},
  year      = {2008},
  doi       = {10.1021/cr068121q},
  url       = {https://doi.org/10.1021/cr068121q}
}

@article{sto_enos_brudz,
  author    = {Osowski, Stanis{\l}aw and Siwek, Krzysztof and Grzywacz, Tomasz and Brudzewski, Kazimierz},
  title     = {Differential Electronic Nose in On-Line Dynamic Measurements},
  journal   = {Metrology and Measurement Systems},
  volume    = {21},
  number    = {4},
  pages     = {649--662},
  year      = {2014},
  doi       = {10.2478/mms-2014-0046},
  url       = {https://doi.org/10.2478/mms-2014-0046}
}

@article{Borowik_enos1,
  author    = {Borowik, Piotr and Adamowicz, Leszek and Tarakowski, Rafa{\l} and Siwek, Krzysztof and Grzywacz, Tomasz},
  title     = {Odor Detection Using an E-Nose with a Reduced Sensor Array},
  journal   = {Sensors},
  volume    = {20},
  number    = {12},
  pages     = {3542},
  year      = {2020},
  doi       = {10.3390/s20123542},
  url       = {https://doi.org/10.3390/s20123542}
}

@article{Jeremie_enos,
  author    = {Loulier, J{\'e}r{\'e}mie and Lefort, Fran{\c{c}}ois and Stocki, Marcin and Asztemborska, Monika and Szmigielski, Rafa{\l} and Siwek, Krzysztof and Grzywacz, Tomasz and Hsiang, Tom and {\'S}lusarski, S{\l}awomir and Oszako, Tomasz and Klisz, Marcin and Tarakowski, Rafa{\l} and Nowakowska, Justyna Anna},
  title     = {Detection of Fungi and Oomycetes by Volatiles Using E-Nose and {SPME}-{GC}/{MS} Platforms},
  journal   = {Molecules},
  volume    = {25},
  number    = {23},
  pages     = {5749},
  year      = {2020},
  doi       = {10.3390/molecules25235749},
  url       = {https://doi.org/10.3390/molecules25235749}
}

@article{Borowik_enos2,
  author    = {Borowik, Piotr and Adamowicz, Leszek and Tarakowski, Rafa{\l} and Siwek, Krzysztof and Grzywacz, Tomasz},
  title     = {On Data Collection Time by an Electronic Nose},
  journal   = {International Journal of Electrical and Computer Engineering (IJECE)},
  volume    = {11},
  number    = {6},
  pages     = {4767--4773},
  year      = {2021},
  doi       = {10.11591/ijece.v11i6.pp4767-4773},
  url       = {https://doi.org/10.11591/ijece.v11i6.pp4767-4773}
}

@article{borowik_zoledzie,
  author    = {Borowik, Piotr and Adamowicz, Leszek and Tarakowski, Rafa{\l} and Wac{\l}awik, Piotr and Oszako, Tomasz and {\'S}lusarski, S{\l}awomir and Tkaczyk, Mi{\l}osz and Stocki, Marcin},
  title     = {Electronic Nose Differentiation between {\textit{Quercus robur}} Acorns Infected by Pathogenic Oomycetes {\textit{Phytophthora plurivora}} and {\textit{Pythium intermedium}}},
  journal   = {Molecules},
  volume    = {26},
  number    = {17},
  pages     = {5272},
  year      = {2021},
  doi       = {10.3390/molecules26175272},
  url       = {https://doi.org/10.3390/molecules26175272}
}

@article{Borowik_ciboria,
  author    = {Borowik, Piotr and Grzywacz, Tomasz and Tarakowski, Rafa{\l} and Tkaczyk, Mi{\l}osz and {\'S}lusarski, S{\l}awomir and Dyshko, Valentyna and Oszako, Tomasz},
  title     = {Development of a Low-Cost Electronic Nose with an Open Sensor Chamber: Application to Detection of {\textit{Ciboria batschiana}}},
  journal   = {Sensors},
  volume    = {23},
  number    = {2},
  pages     = {627},
  year      = {2023},
  doi       = {10.3390/s23020627},
  url       = {https://doi.org/10.3390/s23020627}
}

@article{Hong_array,
  author    = {Zhang, Xijia and Wang, Tao and Ni, Wangze and Zhang, Yongwei and Lv, Wen and Zeng, Min and Yang, Jianhua and Hu, Nantao and Zhan, Rui and Li, Guang and Hong, Zhiqiang and Yang, Zhi},
  title     = {Sensor Array Optimization for the Electronic Nose via Different Deep Learning Methods},
  journal   = {Sensors and Actuators B: Chemical},
  volume    = {418},
  pages     = {136271},
  year      = {2024},
  doi       = {10.1016/j.snb.2024.136271},
  url       = {https://doi.org/10.1016/j.snb.2024.136271}
}

@misc{pen3,
  title        = {{PEN3} Portable Electronic Nose},
  author       = {{AIRSENSE Analytics GmbH}},
  howpublished = {\url{https://airsense.com/en/products/portable-electronic-nose}},
  year         = {n.d.},
  note         = {Accessed: 2025}
}

@misc{Cyranose_320,
  title        = {{Cyranose 320} Electronic Nose},
  author       = {{Sensigent}},
  howpublished = {\url{https://www.sensigent.com/cyranose-320.html}},
  year         = {n.d.},
  note         = {Accessed: 2025}
}

@article{Trung_CMOS,
  author    = {Trung, Tran Quang and Lee, Nae-Eung},
  title     = {Flexible and Stretchable Physical Sensor Integrated Platforms for Wearable Human-Activity Monitoring and Personal Healthcare},
  journal   = {Advanced Materials},
  volume    = {28},
  number    = {22},
  pages     = {4338--4372},
  year      = {2016},
  doi       = {10.1002/adma.201504244},
  url       = {https://doi.org/10.1002/adma.201504244}
}

@article{Cortes_svm,
  author    = {Cortes, Corinna and Vapnik, Vladimir},
  title     = {Support-Vector Networks},
  journal   = {Machine Learning},
  volume    = {20},
  number    = {3},
  pages     = {273--297},
  year      = {1995},
  doi       = {10.1007/BF00994018},
  url       = {https://doi.org/10.1007/BF00994018}
}

@article{Breiman_rf,
  author    = {Breiman, Leo},
  title     = {Random Forests},
  journal   = {Machine Learning},
  volume    = {45},
  number    = {1},
  pages     = {5--32},
  year      = {2001},
  doi       = {10.1023/A:1010933404324},
  url       = {https://doi.org/10.1023/A:1010933404324}
}

@misc{Kingma_adam,
  author        = {Kingma, Diederik P. and Ba, Jimmy},
  title         = {Adam: A Method for Stochastic Optimization},
  year          = {2015},
  eprint        = {1412.6980},
  archivePrefix = {arXiv},
  primaryClass  = {cs.LG},
  doi           = {10.48550/arXiv.1412.6980},
  url           = {https://doi.org/10.48550/arXiv.1412.6980},
  note          = {3rd International Conference on Learning Representations (ICLR)}
}

@book{chapelle2006semi,
  title     = {Semi-Supervised Learning},
  author    = {Chapelle, Olivier and Sch{\"o}lkopf, Bernhard and Zien, Alexander},
  publisher = {The MIT Press},
  year      = {2006},
  doi       = {10.7551/mitpress/9780262033589.001.0001},
  url       = {https://doi.org/10.7551/mitpress/9780262033589.001.0001}
}

@inproceedings{lee_pseudolabel,
  author    = {Lee, Dong-Hyun},
  title     = {Pseudo-Label: The Simple and Efficient Semi-Supervised Learning Method for Deep Neural Networks},
  booktitle = {Workshop on Challenges in Representation Learning, 30th International Conference on Machine Learning (ICML)},
  volume    = {3},
  year      = {2013}
}

@inproceedings{rosenberg2005learning,
  title     = {Semi-Supervised Self-Training of Object Detection Models},
  author    = {Rosenberg, Chuck and Hebert, Martial and Schneiderman, Henry},
  booktitle = {Seventh IEEE Workshop on Applications of Computer Vision (WACV/MOTION)},
  pages     = {29--36},
  year      = {2005},
  doi       = {10.1109/ACVMOT.2005.107},
  url       = {https://doi.org/10.1109/ACVMOT.2005.107}
}

@book{zhou2004ensemble,
  title     = {Ensemble Methods: Foundations and Algorithms},
  author    = {Zhou, Zhi-Hua},
  publisher = {Chapman and Hall/CRC},
  year      = {2012},
  doi       = {10.1201/b12207},
  url       = {https://doi.org/10.1201/b12207}
}

@article{dietterich2000ensemble,
  title     = {An Experimental Comparison of Three Methods for Constructing Ensembles of Decision Trees: Bagging, Boosting, and Randomization},
  author    = {Dietterich, Thomas G.},
  journal   = {Machine Learning},
  volume    = {40},
  number    = {2},
  pages     = {139--157},
  year      = {2000},
  doi       = {10.1023/A:1007607513941},
  url       = {https://doi.org/10.1023/A:1007607513941}
}

@article{rokach2010ensemble,
  title     = {Ensemble-Based Classifiers},
  author    = {Rokach, Lior},
  journal   = {Artificial Intelligence Review},
  volume    = {33},
  number    = {1},
  pages     = {1--39},
  year      = {2010},
  doi       = {10.1007/s10462-009-9124-7},
  url       = {https://doi.org/10.1007/s10462-009-9124-7}
}

@inproceedings{bendale_openmax,
  author    = {Bendale, Abhijit and Boult, Terrance E.},
  title     = {Towards Open Set Deep Networks},
  booktitle = {IEEE Conference on Computer Vision and Pattern Recognition (CVPR)},
  pages     = {1563--1572},
  year      = {2016},
  doi       = {10.1109/CVPR.2016.173},
  url       = {https://doi.org/10.1109/CVPR.2016.173}
}

@inproceedings{liang_odin,
  author        = {Liang, Shiyu and Li, Yixuan and Srikant, R.},
  title         = {Enhancing the Reliability of Out-of-Distribution Image Detection in Neural Networks},
  booktitle     = {6th International Conference on Learning Representations (ICLR)},
  year          = {2018},
  eprint        = {1706.02690},
  archivePrefix = {arXiv},
  doi           = {10.48550/arXiv.1706.02690},
  url           = {https://doi.org/10.48550/arXiv.1706.02690}
}

@article{qu_openset,
  author    = {Qu, Cheng and Liu, Chuanjun and Gu, Yun and Chai, Shuiqin and Feng, Changhao and Chen, Bin},
  title     = {Open-Set Gas Recognition: A Case-Study Based on an Electronic Nose Dataset},
  journal   = {Sensors and Actuators B: Chemical},
  volume    = {360},
  pages     = {131652},
  year      = {2022},
  doi       = {10.1016/j.snb.2022.131652},
  url       = {https://doi.org/10.1016/j.snb.2022.131652}
}

@book{pca_book,
  author    = {Jolliffe, Ian T.},
  title     = {Principal Component Analysis},
  edition   = {2nd},
  publisher = {Springer},
  series    = {Springer Series in Statistics},
  year      = {2002},
  doi       = {10.1007/b98835},
  url       = {https://doi.org/10.1007/b98835}
}

@article{kmeans_book,
  author    = {Lloyd, Stuart P.},
  title     = {Least Squares Quantization in {PCM}},
  journal   = {IEEE Transactions on Information Theory},
  volume    = {28},
  number    = {2},
  pages     = {129--137},
  year      = {1982},
  doi       = {10.1109/TIT.1982.1056489},
  url       = {https://doi.org/10.1109/TIT.1982.1056489}
}

@book{cmeans_book,
  author    = {Bezdek, James C.},
  title     = {Pattern Recognition with Fuzzy Objective Function Algorithms},
  publisher = {Springer},
  year      = {1981},
  doi       = {10.1007/978-1-4757-0450-1},
  url       = {https://doi.org/10.1007/978-1-4757-0450-1}
}

@article{hier_clust,
  author    = {Murtagh, Fionn and Contreras, Pedro},
  title     = {Algorithms for Hierarchical Clustering: An Overview},
  journal   = {WIREs Data Mining and Knowledge Discovery},
  volume    = {2},
  number    = {1},
  pages     = {86--97},
  year      = {2012},
  doi       = {10.1002/widm.53},
  url       = {https://doi.org/10.1002/widm.53}
}

@inproceedings{xie_dec,
  author        = {Xie, Junyuan and Girshick, Ross and Farhadi, Ali},
  title         = {Unsupervised Deep Embedding for Clustering Analysis},
  booktitle     = {33rd International Conference on Machine Learning (ICML)},
  pages         = {478--487},
  year          = {2016},
  eprint        = {1511.06335},
  archivePrefix = {arXiv},
  doi           = {10.48550/arXiv.1511.06335},
  url           = {https://doi.org/10.48550/arXiv.1511.06335}
}

@book{gardner1999electronic,
  title     = {Electronic Noses: Principles and Applications},
  author    = {Gardner, Julian W. and Bartlett, Philip N.},
  publisher = {Oxford University Press},
  year      = {1999},
  doi       = {10.1093/oso/9780198559559.001.0001}
}

@article{loutfi2015review,
  title     = {Electronic Noses for Food Quality: A Review},
  author    = {Loutfi, Amy and Coradeschi, Silvia and Mani, Ganesh Kumar and Shankar, Prabakaran and Rayappan, John Bosco Balaguru},
  journal   = {Journal of Food Engineering},
  volume    = {144},
  pages     = {103--111},
  year      = {2015},
  doi       = {10.1016/j.jfoodeng.2014.07.019},
  url       = {https://doi.org/10.1016/j.jfoodeng.2014.07.019}
}

@article{feng2022fusarium,
  title     = {Early Detection of {Fusarium oxysporum} Infection in Processing Tomatoes ({Solanum lycopersicum}) and Pathogen-Soil Interactions Using a Low-Cost Portable Electronic Nose and Machine Learning Modeling},
  author    = {Feng, Hanyue and Gonzalez Viejo, Claudia and Vaghefi, Niloofar and Taylor, Paul W. J. and Tongson, Eden and Fuentes, Sigfredo},
  journal   = {Sensors},
  volume    = {22},
  number    = {22},
  pages     = {8645},
  year      = {2022},
  doi       = {10.3390/s22228645}
}

@article{peris2013fermentation,
  title     = {On-Line Monitoring of Food Fermentation Processes Using Electronic Noses and Electronic Tongues: A Review},
  author    = {Peris, Miguel and Escuder-Gilabert, Laura},
  journal   = {Analytica Chimica Acta},
  volume    = {804},
  pages     = {29--36},
  year      = {2013},
  doi       = {10.1016/j.aca.2013.09.048}
}

@inproceedings{ICML_Finn,
  author    = {Finn, Chelsea and Abbeel, Pieter and Levine, Sergey},
  title     = {Model-Agnostic Meta-Learning for Fast Adaptation of Deep Networks},
  booktitle = {34th International Conference on Machine Learning (ICML)},
  pages     = {1126--1135},
  year      = {2017},
  doi       = {10.48550/arXiv.1703.03400},
  url       = {https://proceedings.mlr.press/v70/finn17a.html}
}

@inproceedings{Snell_NeurIPS,
  author    = {Snell, Jake and Swersky, Kevin and Zemel, Richard},
  title     = {Prototypical Networks for Few-Shot Learning},
  booktitle = {Advances in Neural Information Processing Systems (NeurIPS)},
  volume    = {30},
  pages     = {4077--4087},
  year      = {2017},
  doi       = {10.48550/arXiv.1703.05175}
}

@article{liu_deepsniffer,
  author    = {Liu, Chuanjun and Miyauchi, Hitoshi and Hayashi, Kenshi},
  title     = {DeepSniffer: A Meta-Learning-Based Chemiresistive Odor Sensor for Recognition and Classification of Aroma Oils},
  journal   = {Sensors and Actuators B: Chemical},
  volume    = {351},
  pages     = {130960},
  year      = {2022},
  doi       = {10.1016/j.snb.2021.130960}
}

@misc{Wanyan_few,
  author        = {Wanyan, Yuyang and Yang, Xiaoshan and Dong, Weiming and Xu, Changsheng},
  title         = {A Comprehensive Review of Few-Shot Action Recognition},
  year          = {2024},
  eprint        = {2407.14744},
  archivePrefix = {arXiv},
  primaryClass  = {cs.CV},
  doi           = {10.48550/arXiv.2407.14744},
  url           = {https://doi.org/10.48550/arXiv.2407.14744}
}

@mastersthesis{Zheng_reinf,
  author    = {Zheng, S.},
  title     = {Deep Reinforcement Learning for Ventilation System Control in Pig Buildings},
  school    = {Universiteit Leiden, Statistics and Data Science},
  year      = {2023}
}

@inproceedings{hendrycks_msp,
  author    = {Hendrycks, Dan and Gimpel, Kevin},
  title     = {A Baseline for Detecting Misclassified and Out-of-Distribution Examples in Neural Networks},
  booktitle = {International Conference on Learning Representations (ICLR)},
  year      = {2017},
  doi       = {10.48550/arXiv.1610.02136}
}

@inproceedings{guo_calibration,
  author    = {Guo, Chuan and Pleiss, Geoff and Sun, Yu and Weinberger, Kilian Q.},
  title     = {On Calibration of Modern Neural Networks},
  booktitle = {Proceedings of the 34th International Conference on Machine Learning (ICML)},
  pages     = {1321--1330},
  year      = {2017},
  doi       = {10.48550/arXiv.1706.04599}
}

@inproceedings{liu_energy,
  author    = {Liu, Weitang and Wang, Xiaoyun and Owens, John D. and Li, Yixuan},
  title     = {Energy-based Out-of-distribution Detection},
  booktitle = {Advances in Neural Information Processing Systems (NeurIPS)},
  volume    = {33},
  pages     = {21464--21475},
  year      = {2020},
  doi       = {10.48550/arXiv.2010.03759}
}

@inproceedings{lee_mahalanobis,
  author    = {Lee, Kimin and Lee, Kibok and Lee, Honglak and Shin, Jinwoo},
  title     = {A Simple Unified Framework for Detecting Out-of-Distribution Samples and Adversarial Attacks},
  booktitle = {Advances in Neural Information Processing Systems (NeurIPS)},
  volume    = {31},
  pages     = {7167--7177},
  year      = {2018},
  doi       = {10.48550/arXiv.1807.03888}
}

@article{vergara_drift,
  author  = {Vergara, Alexander and Vembu, Shankar and Ayhan, Tuba and Ryan, Margaret A. and Homer, Margie L. and Huerta, Ram\'on},
  title   = {Chemical gas sensor drift compensation using classifier ensembles},
  journal = {Sensors and Actuators B: Chemical},
  volume  = {166--167},
  pages   = {320--329},
  year    = {2012},
  doi     = {10.1016/j.snb.2012.01.074}
}

@misc{vergara_dataset,
  author       = {Vergara, Alexander},
  title        = {Gas Sensor Array Drift Dataset},
  howpublished = {UCI Machine Learning Repository},
  year         = {2012},
  doi          = {10.24432/C5RP6W},
  url          = {https://doi.org/10.24432/C5RP6W}
}

@article{lin_platt_svm,
  author  = {Lin, Hsuan-Tien and Lin, Chih-Jen and Weng, Ruby C.},
  title   = {A Note on Platt's Probabilistic Outputs for Support Vector Machines},
  journal = {Machine Learning},
  volume  = {68},
  number  = {3},
  pages   = {267--276},
  year    = {2007},
  doi     = {10.1007/s10994-007-5018-6}
}

@article{dean_top10fungal,
  author    = {Dean, Ralph and van Kan, Jan A. L. and Pretorius, Zacharias A. and Hammond-Kosack, Kim E. and Di Pietro, Antonio and Spanu, Pietro D. and Rudd, Jason J. and Dickman, Martin and Kahmann, Regine and Ellis, Jeff and Foster, Gary D.},
  title     = {The Top 10 Fungal Pathogens in Molecular Plant Pathology},
  journal   = {Molecular Plant Pathology},
  volume    = {13},
  number    = {4},
  pages     = {414--430},
  year      = {2012},
  doi       = {10.1111/j.1364-3703.2011.00783.x}
}

@article{mohareb2016ensemble,
  author    = {Mohareb, Fady and Papadopoulou, Olga and Panagou, Efstathios and Nychas, George-John and Bessant, Conrad},
  title     = {Ensemble-Based Support Vector Machine Classifiers as an Efficient Tool for Quality Assessment of Beef Fillets from Electronic Nose Data},
  journal   = {Analytical Methods},
  volume    = {8},
  number    = {18},
  pages     = {3711--3721},
  year      = {2016},
  doi       = {10.1039/C6AY00147E},
  url       = {https://doi.org/10.1039/C6AY00147E}
}

\end{document}